\documentclass[twocolumn]{aastex63}

\usepackage{amsmath}

\newcommand{\thi}{$\tau_{\rm HI}(v)$}
\newcommand{\hi}{H\textsc{i}}
\newcommand{\nhi}{$N(\rm H\textsc{i})$}
\newcommand{\nhithin}{$N(\rm H\textsc{i})^*$}
\newcommand{\rhi}{$\mathcal{R}_{\rm HI}$}
\newcommand{\fcnm}{$f_{\rm CNM}$}
\newcommand{\ebv}{$E(B-V)$}
\newcommand{\R}{$\mathcal{R}$}
\newcommand{\nabs}{$58$}
\newcommand{\tbexp}{$T_{B,\rm exp}(v)$}
\newcommand{\nsyn}{$38781$}
\newcommand{\nval}{$11634$}

\graphicspath{{./}{figures/}}

\received{\today}
\submitjournal{ApJ}

\shorttitle{A CNN for the CNM}
\shortauthors{Murray, Peek, \& Kim}

\begin{document}

\title{Extracting the cold neutral medium from HI emission with deep learning: Implications for Galactic foregrounds at high latitude}

\correspondingauthor{C.\,E.\,M.}
\email{clairemurray56@gmail.com}

\author[0000-0002-7743-8129]{Claire E. Murray}
\altaffiliation{NSF Astronomy \& Astrophysics Postdoctoral Fellow}
\affil{Department of Physics \& Astronomy, 
Johns Hopkins University,
3400 N. Charles Street, 
Baltimore, MD 21218}

\author[0000-0003-4797-7030]{J. E. G. Peek}
\affil{Space Telescope Science Institute,
3700 San Martin Drive, 
Baltimore, MD, 21218}
\affil{Department of Physics \& Astronomy, 
Johns Hopkins University,
3400 N. Charles Street, 
Baltimore, MD 21218}

\author[0000-0003-2896-3725]{Chang-Goo Kim}
\affil{Department of Astrophysical Sciences, Princeton University, Princeton, NJ 08544, USA}
\affil{Center for Computational Astrophysics, Flatiron Institute, New York, NY 10010, USA}

\begin{abstract}
Resolving the phase structure of neutral hydrogen (\hi) is crucial for understanding the life cycle of the interstellar medium (ISM). However, accurate measurements of \hi\ temperature and density are limited by the availability of background continuum sources for measuring \hi\ absorption. Here we test the use of deep learning for extracting \hi\ properties over large areas without optical depth information. We train a 1D convolutional neural network using synthetic observations of 3D numerical simulations of the ISM to predict the fraction of cold neutral medium (\fcnm) and the correction to the optically-thin \hi\ column density for optical depth (\rhi) from $21\rm\,cm$ emission alone. We restrict our analysis to high Galactic latitudes ($|b|>30^{\circ}$), where the complexity of spectral line profiles is minimized. We verify that the network accurately predicts \fcnm\ and \rhi\ by comparing the results with direct constraints from $21\rm\,cm$ absorption. By applying the network to the GALFA-\hi\ survey, we generate large-area maps of \fcnm\ and \rhi. Although the overall contribution to the total \hi\ column of cold neutral medium (CNM)-rich structures is small ($\sim 5\%$), we find that these structures are ubiquitous. Our results are consistent with the picture that small-scale structures observed in $21\rm\,cm$ emission aligned with the magnetic field are dominated by CNM. Finally, we demonstrate that the observed correlation between \hi\ column density and dust reddening (\ebv) declines with increasing \rhi, indicating that future efforts to quantify foreground Galactic \ebv\ using \hi, even at high latitudes, should increase fidelity by accounting for \hi\ phase structure.
\end{abstract}


\section{Introduction} \label{sec:intro}

Neutral atomic hydrogen (\hi) plays a fundamental role in the evolutionary life cycle of galaxies. \hi\ provides the fuel reservoir from which star-forming molecular clouds form \citep[e.g.,][]{clark2012, klessen2016}, and also provides an important source of radiation shielding for the formation and survival of interstellar molecules and dust \citep[e.g.,][]{sternberg2014, lee2015}.

\begin{figure*}
\begin{center}
\includegraphics[width=0.98\textwidth]{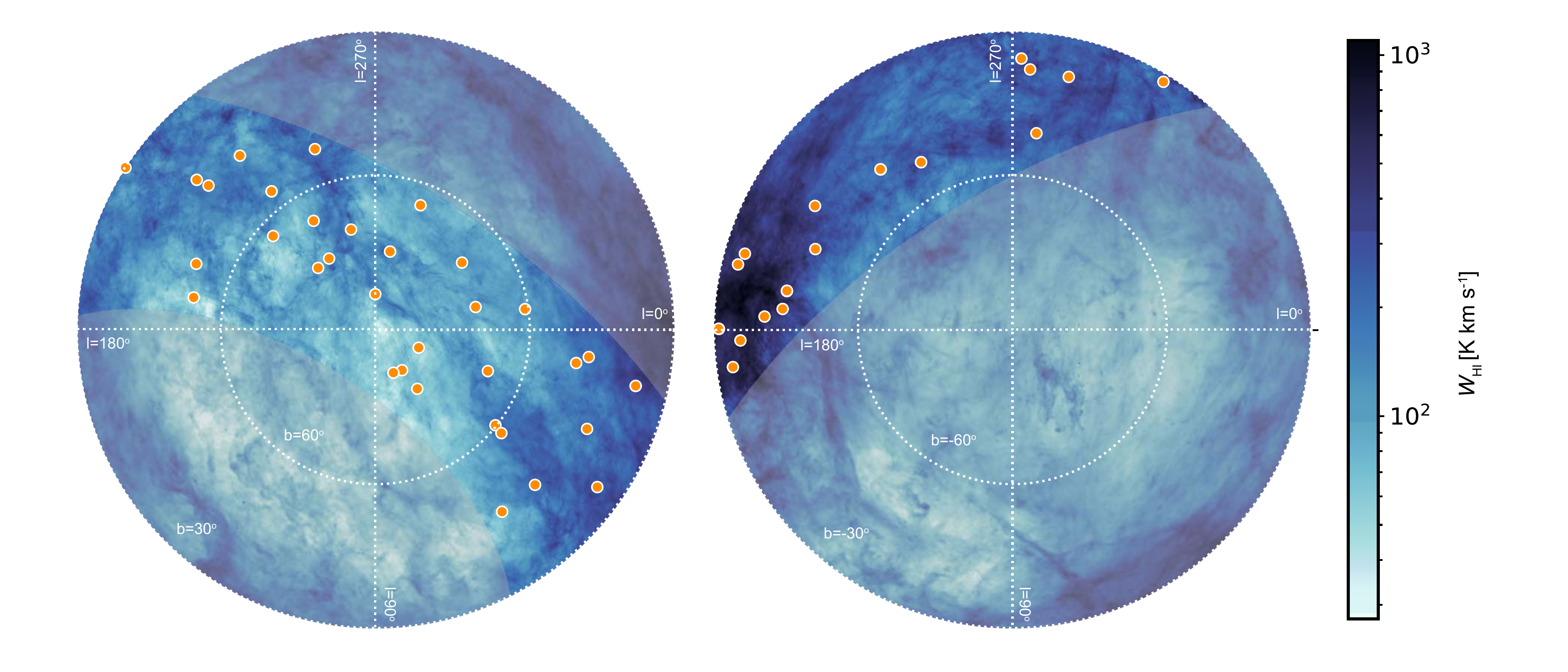}
\caption{Maps of integrated $21\rm\,cm$ brightness temperature ($W(\rm HI)$) from the GALFA-\hi\ survey \citep{peek2018} and the HI4PI survey \citep{hi4pi2016} (shaded) in zenith equal area (ZEA) projection (left: north; right: south) for $|b|>30^{\circ}$. The coordinates of the \nabs\ LOS along which we have obtained sensitive \thi\ observations are overlaid (orange circles).}
\label{f:abs_sources}
\end{center}
\end{figure*}

In thermal equilibrium between dominant sources of heating and cooling in the ISM, two thermally stable phases of \hi\ emerge: the cold neutral medium (CNM) and the warm neutral medium (WNM), with kinetic temperature and density of ($T_k, n$) = ($60-260\rm\,K$, $7-70\rm\,cm^{-3}$) and ($T_k, n$) = ($5000-8300\rm\,K$, $0.2-0.9\rm\,cm^{-3}$) respectively \citep{field1969, mckee1977, wolfire2003}. The WNM dominates the mass budget of \hi\ in the ISM, comprising roughly $\sim 50\%$, whereas the CNM accounts for $30\%$ of the total \hi\ mass. The remaining $\sim 20\%$ of \hi\ exists in a thermally unstable phase with intermediate temperature and density \citep{heiles2003, murray2018b}. Distinguishing these phases (CNM, WNM, unstable medium) and how mass is transferred between them is fundamental for understanding how star-forming clouds form and evolve within the ambient ISM. 

In addition, quantifying the effects of dust grains mixed within all \hi\ phases is crucial for measuring accurate color and brightness of extragalactic objects. Empirical correlations between \hi\ emission at $21\rm\,cm$ and dust emission in the infrared \citep[e.g.,][]{low1984, boulanger1996, burstein1982, lenz2017} or dust extinction \citep[e.g.,][]{sturch1969, bohlin1978} 
indicate that dust and gas are well-mixed. However, the magnitude of this correlation depends on the mixture of ISM phases along the line of sight \citep[i.e., ionized, atomic, molecular;][]{liszt2014, lenz2017, nguyen2018}. Quantifying these variations is essential for understanding how dust grains evolve in disparate ISM conditions, and also for calibrating and constructing precise foreground reddening maps in aid of cosmological studies \citep{lenz2017}. In particular, \hi\ emission is a promising tracer of foreground reddening, as it is unaffected by contamination from emission by the very same targets in need of de-reddening \citep{chiang2019}.

However, measuring accurate \hi\ temperature and density via the $21\rm\,cm$ line requires measurements of both emission and absorption, and absorption observations are limited by the availability of background sources of continuum radiation. For example, although currently-available observations of $21\rm\,cm$ absorption have constrained global properties of the local ISM, including the temperature, column density and mass fractions of CNM, WNM and unstable gas \citep[e.g.,][]{dickey1978,crovisier1978,mebold1982,heiles2003,roy2013,murray2014,murray2015,murray2018a,murray2018b}, they are too sparse for resolving the spatial distributions of these disparate phases. Ongoing surveys \citep[e.g.,][]{dickey2013} will alleviate this problem by increasing the number of sightlines by orders of magnitude, but even these will not fully sample the local ISM.

\begin{figure*}
\begin{center}
\includegraphics[width=\textwidth]{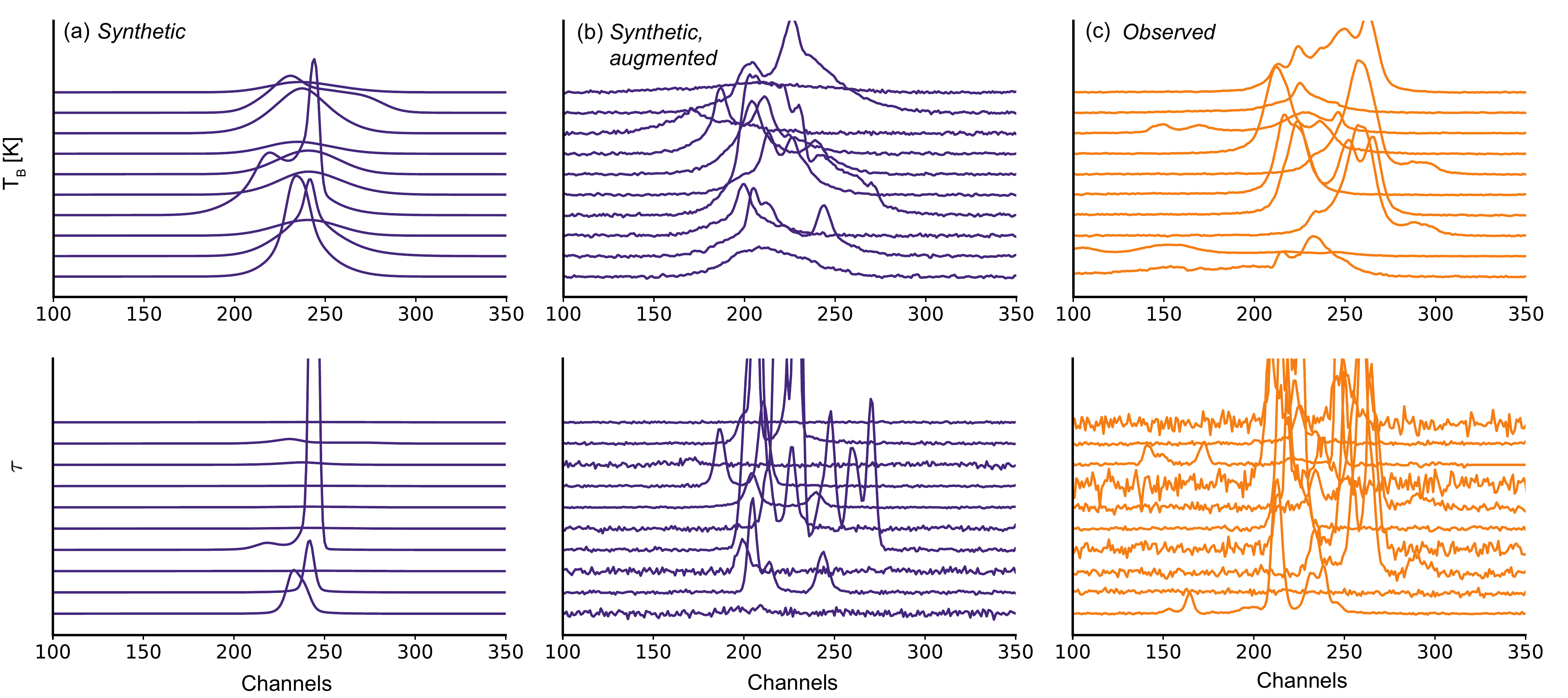}
\caption{Synthetic $21\rm\,cm$ brightness temperature ($T_B$; top row) and optical depth ($\tau$; bottom row) observations from numerical simulations (KOK13, KOK14) before (panel a) and after (panel b) augmentation (Section~\ref{sec:synthetic_spectra}), compared with real observations (panel c) for 15 randomly-selected LOS. The spectra within each panel are offset to illustrate the line shapes and noise properties, and the panels in each row are plotted with the same, arbitrary scales. }
\label{f:example_spectra}
\end{center}
\end{figure*}

Faced with this reality, indirect estimates for \hi\ properties based on the velocity structure of $21\rm\,cm$ emission have been used to diagnose \hi\ properties across local and extragalactic regimes. A common method for constructing large-area maps of \hi\ properties is to decompose $21\rm\,cm$ emission spectra into Gaussian functions and infer the temperature and column density of each feature based on its spectral line width and amplitude \citep[e.g.,][]{matthews1957,takakubo1966,mebold1972,haud2007,kalberla2018}. Unfortunately, Gaussian decomposition is complicated by strong blending of lines in velocity, non-Gaussian line shapes and baseline structure systematics, suggesting it is not a valid approach \citep{dickey1990}. However, sophisticated tools have tackled these problems by imposing continuity for extracting coherent structures and automating component selection \citep[e.g.,][]{marchal2019,riener2020}.

In this work we will test the use of deep learning for diagnosing \hi\ properties. Specifically, we will apply a convolutional neural network (CNN) to $21\rm\,cm$ emission to infer properties of \hi\ which formally require $21\rm\,cm$ absorption information. This approach is inspired by the application of CNNs for extracting stellar properties from spectroscopic surveys \citep{bailerjones1997,bailerjones2000,manteiga2010,fabbro2018}. The availability of large, high-resolution observational surveys and new, increasingly-realistic simulations of the ISM for training have made this possible only recently. The advantage of the approach over previous methods is that the analysis is efficient and reproducible (i.e., requiring no subjective input), enabling the analysis of synthetic data with prohibitively large sizes for human input, as well as objective comparisons between observed and simulated data. 

This paper is organized as follows. In Section~\ref{sec:data} we introduce the real and synthetic \hi\ observations used in our analysis, as well as the relevant \hi\ properties under consideration. In Section~\ref{sec:CNN} we introduce and explain the construction of the CNN, including its architecture and training. In Section~\ref{sec:results} we present the results of applying the trained CNN to the \hi\ observations. Finally, in Section~\ref{sec:discussion} we discuss the results in the context of the local ISM. 

\section{Data}
\label{sec:data}

In the following section we describe the data used in our analysis. First, we build a sample of observed $21\rm\,cm$ spectral line pairs probing gas in the local ISM. Next, we build a sample of synthetic $21\rm\,cm$ spectral line pairs which will be used to train our CNN to analyze the real observations. Finally, we describe the relevant properties of the CNM which we will analyze in this work. 

\subsection{21cm Emission}

To trace \hi\ emission throughout the local ISM, we use the Galactic Arecibo L-band Feed Array Survey \citep[GALFA-HI;][]{peek2011,peek2018} at the Arecibo Observatory. GALFA-\hi\ is the highest angular resolution ($\sim4^\prime$), highest spectral resolution ($0.18\rm\,km\,s^{-1}$) large-area ($13,000\,\rm deg^{2}$) Galactic $21\rm\,cm$ emission survey to date. From the GALFA-\hi\ DR2 data release \citep{peek2018} we select the ``Narrow" data cubes of \hi\ brightness temperature ($T_B(v)$) and extract all Galactic velocities (defined as velocity in the Local Standard of Rest (LSR) $|v_{\rm LSR}|<90\rm\,km\,s^{-1}$)\footnote{All velocities quoted in this work (``$v$") are in the LSR frame.}. For visualization purposes, where GALFA-\hi\ is unavailable (i.e., outside of the Arecibo field of view), we use data from the HI4PI survey \citep{hi4pi2016}, which has lower angular resolution ($16^\prime.2$) and velocity resolution ($1.3\rm\,km\,s^{-1}$ per channel) but covers the full sky.\footnote{In this work we will focus on analyzing GALFA-\hi\ to test the validity of CNNs, and will include HI4PI data in future work to investigate the effects of angular and velocity resolution on the extraction of CNM properties with deep learning methods.}

In Figure~\ref{f:abs_sources} we display two zenith-equal-area (ZEA) projection maps of integrated brightness temperature ($W({\rm HI}) = \int T_B(v)\, dv$) from GALFA-\hi\ and HI4PI.

\subsection{21 cm Absorption}

As a sensitive probe of the absorption properties of \hi\ in the local ISM, we assemble a sample of available $21\rm\,cm$ optical depth spectra (\thi) from the literature. We use targeted surveys of \hi\ absorption focused outside of the Galactic plane with publicly available data, to avoid the complex spectral line structures observed at the lowest Galactic latitudes which include significant saturation and self-absorption effects. For this study, we restrict our analysis to the GALFA-\hi\ footprint: $0<\alpha_{2000}<360^{\circ}$, $1<\delta_{2000}<38^{\circ}$ (lowest and highest declinations excluded due to systematic artifacts), and $|b|>30^{\circ}$.

\begin{figure}
\begin{center}
\includegraphics[width=0.48\textwidth]{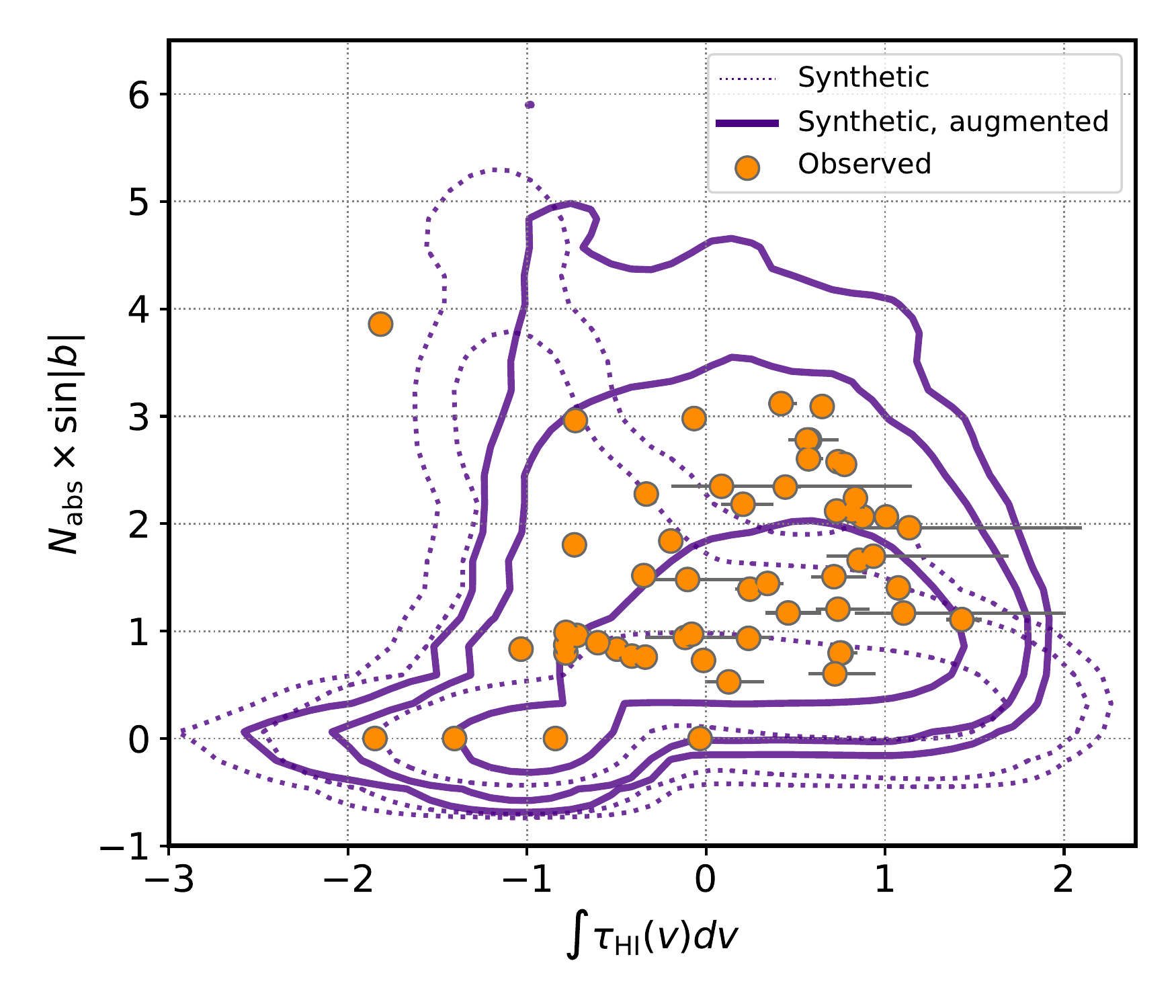}
\caption{Comparing the number of absorption lines per path length ($N_{\rm abs}\times \sin(|b|)$) as a function of the integrated \hi\ optical depth along the line of sight ($\int \tau_{\rm HI}(v)$) for the synthetic $21\rm\,cm$ data set before (dotted purple) and after (solid purple) augmentation (Section~\ref{sec:aug}; contours represent $68,\,95,$ and $99\%$ of each distribution) and for the observed sources (orange circles). The augmented synthetic sample agrees much better with the observed sample than the original synthetic sample.}
\label{f:nabs_comparison}
\end{center}
\end{figure}

\begin{figure*}
\begin{center}
\includegraphics[width=\textwidth]{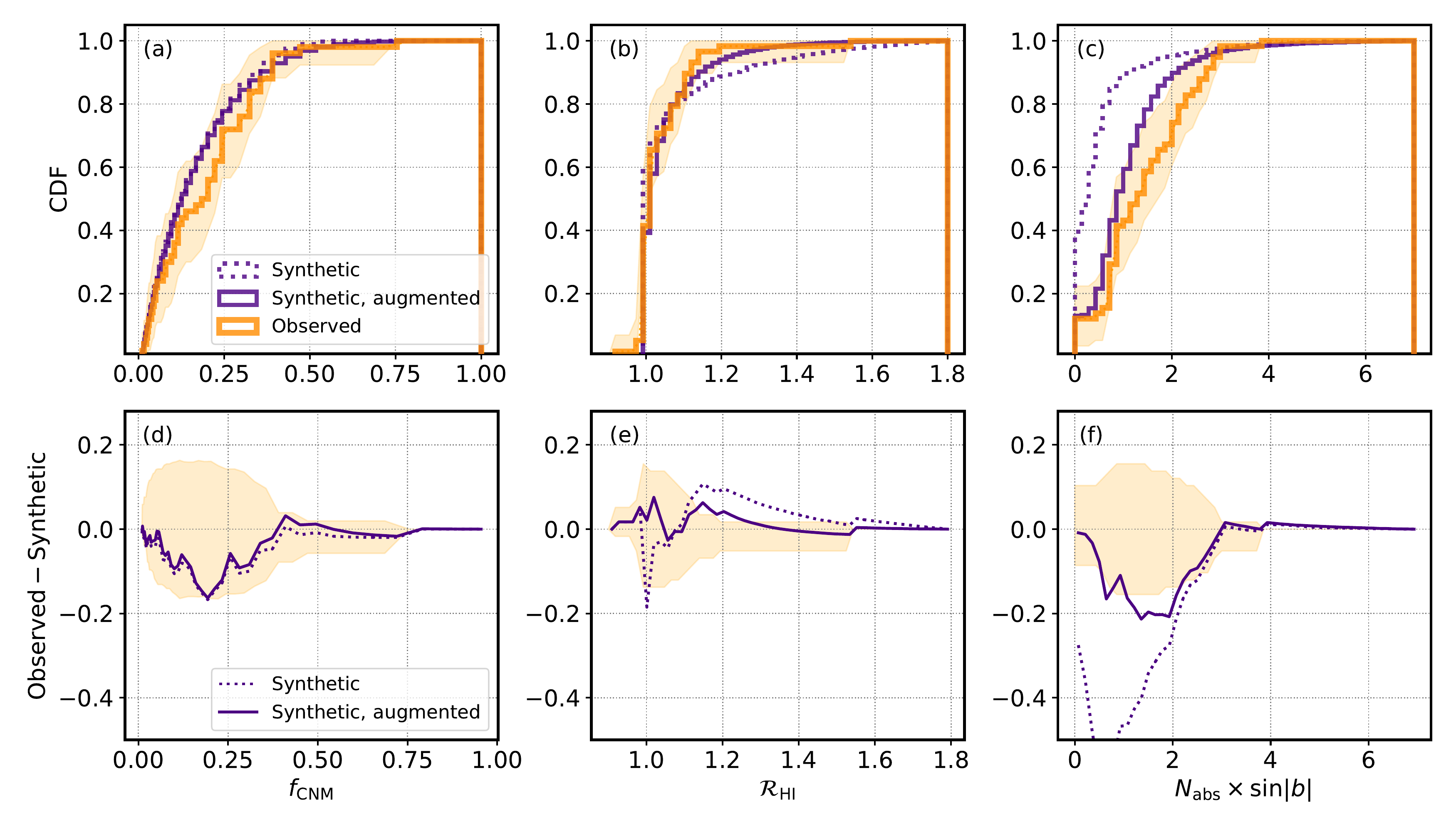}
\caption{Cumulative distribution functions of \fcnm\ (a; Equation~\ref{e:fcnm}); \rhi\ (b; Equation~\ref{e:rhi}) and the number of absorption lines per path length (c; $N_{\rm abs}\times \sin{|b|}$) for the \nabs\ observed (orange) and synthetic $21\rm\,cm$ spectral line pairs (purple). For the synthetic data, include the original sample (dotted) and the augmented sample (solid). The uncertainties on the observed distributions are computed by bootstrapping the \nabs\ LOS sample. In panels (d), (e) and (f) we include differential distributions (observed-synthetic), including the uncertainty envelopes from the observed distributions.}
\label{f:properties}
\end{center}
\end{figure*}

\begin{enumerate}
\item \noindent \emph{21-SPONGE}: From the $21\rm\,cm$ Spectral Line Observations of Neutral Gas with the Karl G. Jansky Very Large Array (VLA) survey \citep[21-SPONGE;][]{murray2015,murray2018b} we select the $30$ spectra in our region of interest. 21-SPONGE is the most sensitive survey for \thi\ to date at the VLA, and achieved excellent optical depth sensitivity and velocity resolution (median root mean square (RMS) uncertainty in \thi\ of $\sigma_{\tau \rm HI}\sim0.001$ per $0.42\rm\,km\,s^{-1}$ channels).

\item \noindent \emph{Millennium Survey}: In addition to high-latitude spectra from 21-SPONGE, we include \thi\ spectra from the Millennium Arecibo $21\rm\,cm$ Absorption-Line Survey \citep{heiles2003, heiles2003b}. These spectra have lower optical depth sensitivity than 21-SPONGE ($\sigma_{\tau \rm HI} = 0.01$ per $0.18\rm\,km\,s^{-1}$ channels).
We select the $28$ spectra which are unique relative to the 21-SPONGE sample in our region of interest which do not show spurious spectral artifacts (i.e., ``negative absorption" from resolved \hi\ emission, significant baseline artifacts, defined as exceeding $\pm 3\sigma$). 
\end{enumerate}

In Figure~\ref{f:abs_sources}, we include the positions of the \nabs\ high-latitude absorption targets gathered from 21-SPONGE and the Millennium Survey. The source names and their coordinates are also included in Table~\ref{tab:sources} (Appendix~\ref{ap:sources}). 

\subsection{Sample construction}
\label{sec:sample}

Given the \hi\ observations, we next construct a sample of emission and absorption spectral line pairs. First, we smooth and re-grid all GALFA-\hi\ cubes and all \thi\ spectra to $0.42\rm\,km\,s^{-1}$ per channel resolution (i.e., the limiting channel resolution of 21-SPONGE). Unfortunately, the presence of each radio continuum source precludes us from extracting $21\rm\,cm$ brightness temperature spectra which sample exactly the same \hi\ as the absorption observations. To simulate the ``expected" emission spectrum (\tbexp) in the absence of the continuum source, we extract spectra from all pixels within a $9\times9$ pixel grid surrounding each source, remove the innermost $3\times3$ pixels which are contaminated by absorption due to the source, and average the remaining 72 spectra. This method was shown by \citet{lee2015} to agree within $10\%$ with the more sophisticated interpolation method employed by \citet{heiles2003} to estimate the expected brightness temperature profile. 

Next, as the brightness temperature of \hi\ contributes significantly to the system temperature of a radio receiver at Galactic velocities, the uncertainty in \tbexp\ and \thi\ depends on velocity. For each spectrum, we estimate the uncertainty \tbexp\ and \thi\ as a function of velocity ($\sigma_{T_B,\rm exp}(v)$, $\sigma_{\tau \rm HI}(v)$) following the methods described in \citet[][Section 3.2]{murray2015}, which were applied following \citet{roy2013}. 

As a final step, we restrict each spectral pair to LSR velocities $|v_{\rm LSR}|<90 \rm\,km\,s^{-1}$, where the bulk of Milky Way \hi\ emission lies. Although the $30$ 21-SPONGE \thi\ spectra only cover $\sim \pm 50\rm\,km\,s^{-1}$ in velocity, we inspect each sightline from this sample and observe no significant evidence for emission from high-velocity structures which might have significant absorption with $50{\rm\,km\,s^{-1}}<|v|<90{\rm\,km\,s^{-1}}$ which would be missed by 21-SPONGE. For those 21-SPONGE channels without observed optical depth, we generate synthetic Gaussian noise with RMS equal to $\sigma_{\tau \rm HI}$.

The result is a sample of \nabs\ spectral line pairs (\thi\ and \tbexp) and their uncertainties ($\sigma_{\tau \rm HI}(v)$ and $\sigma_{T_B,\rm exp}(v)$).

\subsection{Synthetic \hi\ Spectra}
\label{sec:synthetic_spectra}

To generate a training set, we start with the library of synthetic $21\rm\,cm$ emission and absorption spectral line pairs presented by \citet[][hereafter KOK14]{kim2014}. These spectra were extracted from the three-dimensional hydrodynamic simulations of the Milky Way ISM by \citet[][hereafter KOK13]{kim2013}. We refer the reader to KOK13 for a full description of the simulations, which include supernova feedback, time-varying heating and cooling of the ISM, galactic differential rotation, self-gravity from gas and external gravity from stars and dark matter.  

To construct the synthetic spectra, KOK14 placed an ``observer" in the center of the simulations and extracted salient \hi\ properties (temperature, density, velocity) as function of path length along the line of sight (LOS) from $10^4$ random positions in Galactic latitude and longitude ($|b|>5^{\circ}$). We select the simulation (denoted ``QA10") with galactic rotation applied assuming an angular velocity of $\Omega = 28 \rm\,km \,s^{-1}$ and gas surface density $\Sigma = 10\,\rm M_{\odot}\, pc^{-2}$ (KOK13). From the \hi\ properties extracted from this simulation, KOK14 applied analytical radiative transfer and line excitation (c.f., Section 2.3; KOK14) to estimate $21\rm\,cm$ emission and absorption as a function of velocity along each LOS. 

\subsubsection{Augmentation}
\label{sec:aug}

From the standpoint of integrated properties of the local ISM, the synthetic $21\rm\,cm$ spectra generally agree with the results of all-sky surveys (KOK14). However, detailed comparisons between the synthetic $21\rm\,cm$ velocity structure and 21-SPONGE reveal key differences, including the fact that KOK14 spectra feature fewer distinct velocity components than observed \thi\ spectra \citep{murray2017}. 

To improve this comparison, we augment the synthetic spectra. For each synthetic LOS, we generate four new synthetic spectral line pairs by adding (respectively) 2, 3, 4 and 5 randomly-selected spectra from the sample, each of which are modified by random velocity shifts (selected from a uniform distribution of velocities between $\pm20\rm\,km\,s^{-1}$) or being randomly flipped in velocity across $v=0\rm\,km\,s^{-1}$. 
The result is a sample of $4\times10^4$ synthetic $21\rm\,cm$ spectral line pairs ($T_{B,\rm syn}(v)$, $\tau_{\rm HI, syn}(v)$).
We re-sample the synthetic pairs to $0.42\rm\,km\,s^{-1}$ per channel resolution to match the observed sample.

Next, we introduce realistic noise properties. For $T_{B,\rm syn}(v)$, to directly incorporate noise from instrumental effects and observing conditions from GALFA-\hi, we randomly extract signal-free velocity windows from outside the Galactic velocity range used in our analysis (i.e., $|v|\geq90\rm\,km\,s^{-1}$). We then add the noise-only spectra to the augmented $T_{B,\rm syn}(v)$ samples, discarding the resulting spectra with invalid channel values (i.e., NaNs, which comprise $3\%$ of the sample). For $\tau_{\rm HI,syn}(v)$, we add Gaussian noise with RMS amplitude selected randomly from the distribution of $\sigma_{\tau}$ in the observed sample. After adding the noise, we generate the uncertainty arrays ($\sigma_{\tau, \rm HI, syn}(v)$ and $\sigma_{T_{B,\rm syn}}(v)$) for each pair in the same manner as done for the observed \thi\ and \tbexp\ spectra (Section~\ref{sec:sample}), to simulate the effect of increased system temperature at line center due to the contribution of $21\rm\,cm$ emission.\footnote{We note that $T_{B,\rm syn}(v)$ were not generated using the same method of averaging neighboring spectra (as done for the observed $T_{B,\rm exp}(v)$ to simulate the expected emission in the absence of a continuum source), which presents an additional contribution to the uncertainty. We plan to address this effect directly in future analysis of synthetic data cubes from next-generation simulations.} Finally, as done for the observed sample, we restrict each spectral pair to velocities $\pm 90\rm\,km\,s^{-1}$.

The result is a sample of \nsyn\ synthetic $21\rm\,cm$ spectral line pairs ($\tau_{\rm HI, syn}(v)$ and $T_{B,\rm syn}(v)$) and their uncertainties ($\sigma_{\tau, \rm HI, syn}(v)$ and $\sigma_{T_{B,\rm syn}}(v)$) at the re-sampled GALFA-\hi\ velocity resolution ($0.42\rm\,km\,s^{-1}$ per channel). In Figure~\ref{f:example_spectra} we illustrate the augmentation process by comparing the original synthetic spectral pairs (a) with the augmented synthetic pairs (b) and the observed pairs (c). By eye, the augmented spectral pairs agree much better with the observations.

Beyond inspection, to verify that the augmentation process produced realistic synthetic spectral lines, we compare the number of absorption components between the synthetic and observed samples. To count the number of absorption components for each observed and synthetic LOS, we smooth each $\tau_{\rm HI}(v)$ spectrum with a Gaussian kernel ($\sigma=2$ channels) to suppress spurious noise spikes and compute the derivative. Components are defined as peaks (locations of negative curvature) with significant amplitude (defined by $>3\sigma_{T_{\rm B}}(v)$). To compute the number of components per path length, we multiply the total number ($N_{\rm abs}$) by $\sin{|b|}$ to approximate the number of components in the vertical direction. For each synthetic LOS, we use the average $|b|$ of the spectra comprising the augmented LOS.  

In Figure~\ref{f:nabs_comparison} we compare the number of absorption components per path length with the integrated optical depth along the LOS ($\int \tau_{\rm HI}(v)dv$). In addition to the augmented and observed samples, we include results from the original synthetic spectral line set, to which we added realistic noise but not randomly shifted or flipped spectra. We observe that although the original synthetic sample features significantly fewer components per path length than the observations \citep[i.e., in agreement with analysis by][]{murray2017}, the augmented synthetic sample compares much more favorably. 

\subsection{\hi\ Properties}

Armed with observed and synthetic $21\rm\,cm$ spectral line pairs and their uncertainties, we compute salient \hi\ properties. In this work, we will consider properties which reflect the balance of \hi\ phases (WNM, CNM) and the thermodynamic state of the gas along the line of sight, including the contribution of optically-thick \hi\ to the total \hi\ column density (\rhi) and the mass fraction of the CNM (\fcnm).

\subsubsection{Column Density Correction Factor}

For \hi\ with optical depth \thi\ and excitation (a.k.a., ``spin") temperature $T_s(v)$, the \nhi\ is given by,

\begin{equation}
N({\rm HI}) = C_0 \,\int \tau_{\rm HI}\,\, T_s\,\, dv , 
\label{e:nhi}
\end{equation}

\noindent where $C_0 = 1.823 \times10^{18}\rm\,cm^{-2}/(K\,km\,s^{-1})$ \citep[e.g.,][]{draine2011}. In the isothermal approximation
(i.e., assuming each velocity channel is dominated by a single temperature component), the spin temperature at a given velocity channel can be approximated as 

\begin{equation}
    T_s(v) = \frac{T_B(v)}{1-e^{-\tau_{\rm HI}(v)}} ,
    \label{e:ts}
\end{equation}

\noindent which yields

\begin{equation}
N({\rm HI}) \simeq C_0 \int \frac{\tau_{\rm HI} \,\,T_B}{(1-e^{-\tau_{\rm HI}})}\,\, dv ,
\label{e:nhi_iso}
\end{equation}

\noindent \citep[e.g.,][]{dickey1982}. In the optically-thin limit ($\tau_{\rm HI}\ll 1$), Equation~\ref{e:nhi_iso} reduces to,

\begin{equation}
N({\rm HI})^* = C_0\,\int T_B \, dv . 
\label{e:nhi_thin}
\end{equation}

\begin{figure}
\begin{center}
\includegraphics[width=0.45\textwidth]{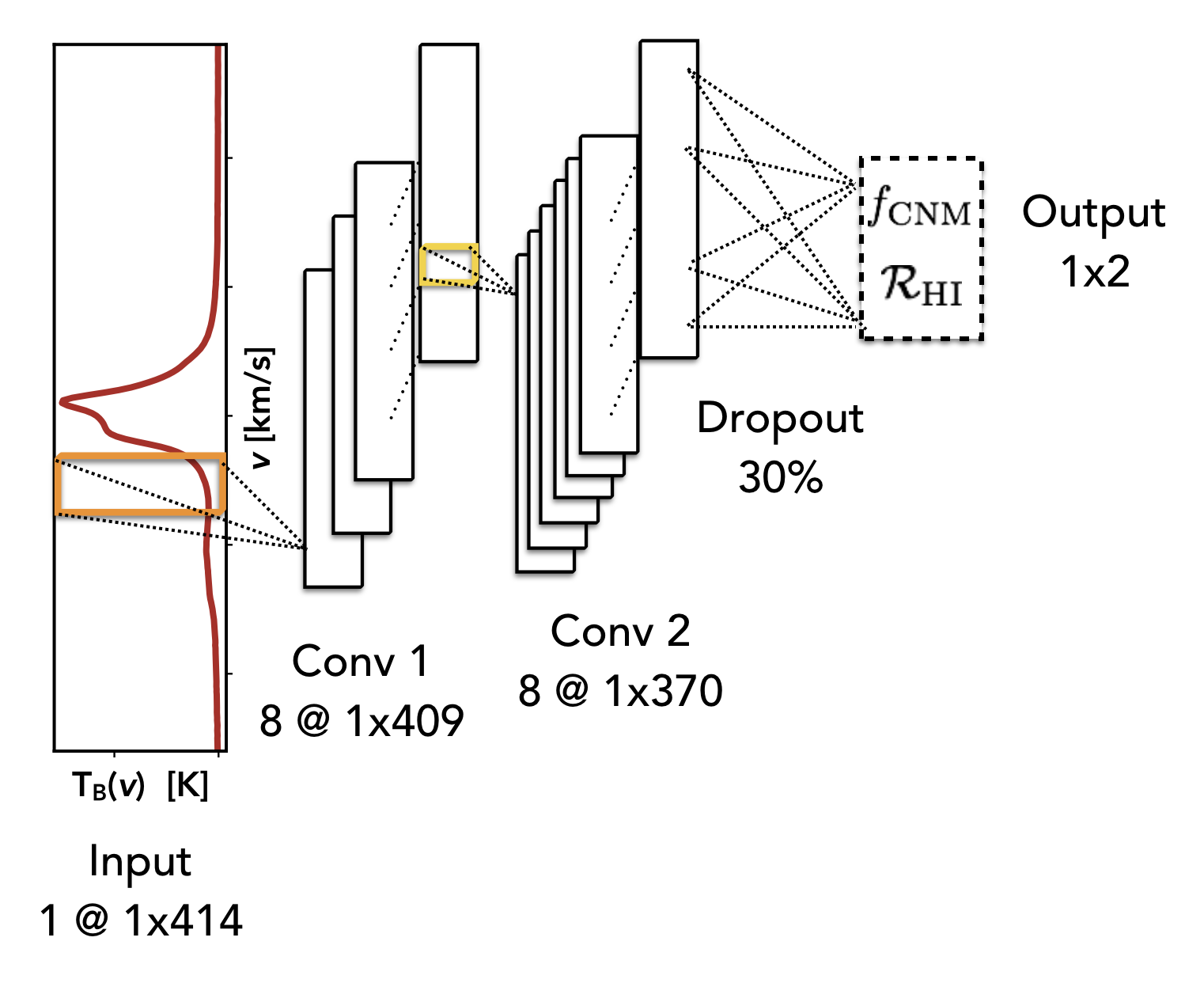}
\caption{Illustration of CNN architecture, including an input layer (e.g,. $T_{B,\rm syn}(v)$, $1\times 414$ channels) two convolutional layers with 8 filters each (convolutional windows of size $6$ and $40$ channels respectively), a dropout of $30\%$ and a final fully-connected output layer ($1\times 2$) containing the classes (\rhi\ and \fcnm).}
\label{f:architecture}
\end{center}
\end{figure}

\noindent In the absence of optical depth information, Equation~\ref{e:nhi_thin} is used to approximate \nhi. Clearly, in the presence of significant \thi, \nhithin\ will significantly underestimate \nhi. By comparing these two estimates, optical depth information can be deduced. It is noteworthy that \nhi\ is itself an approximate estimate of the true column density.

Sensitive surveys for \thi\ in the local ISM \citep{stanimirovic2014, lee2015, murray2018b} have shown that the isothermal approximation to \nhi\ (Equation~\ref{e:nhi_iso}) is consistent with the results of detailed decomposition of multi-phase components with distinct densities and temperatures, especially for low column densities (\nhi$<5\times10^{20}\rm\,cm^{-2}$). This is also consistent with the results of numerical simulations of the Galactic ISM, which find that synthetically-observed column densities in the isothermal limit (i.e., Equation~\ref{e:nhi_iso}) agree within $5\%$ of the true LOS column density \citep{kim2014}. To minimize the complexity of the \nhi\ computation, and considering that we restrict our analysis to high Galactic latitudes, we will therefore use Equation~\ref{e:nhi_iso} to compute \nhi.

To quantify the contribution of optically-thick \hi, we compute \rhi, the ratio between \nhi\ and \nhithin,

\begin{equation}
    \mathcal{R}_{\rm HI} = \frac{N({\rm HI})}{N({\rm HI})^*}.
    \label{e:rhi}
\end{equation}


\subsubsection{Fraction of Cold Neutral Medium}

To compute \fcnm, a standard approach is to extract the spectral properties of individual \hi\ structures along each LOS and estimate their unique properties, including temperature and density \citep[e.g.,][]{murray2018b}. The value of \fcnm\ is then computed from the sum of \nhi\ in the CNM phase (e.g., $T_s<350\rm\,K$) relative to the total \nhi\ along the LOS. Despite considerable uncertainty in this approach due to the presence of strongly-blended line profiles, the resulting \fcnm\ values are consistent with the approximation of a single CNM temperature along the full LOS \citep{dickey2000}, given by,

\begin{figure*}
\begin{center}
\includegraphics[width=\textwidth]{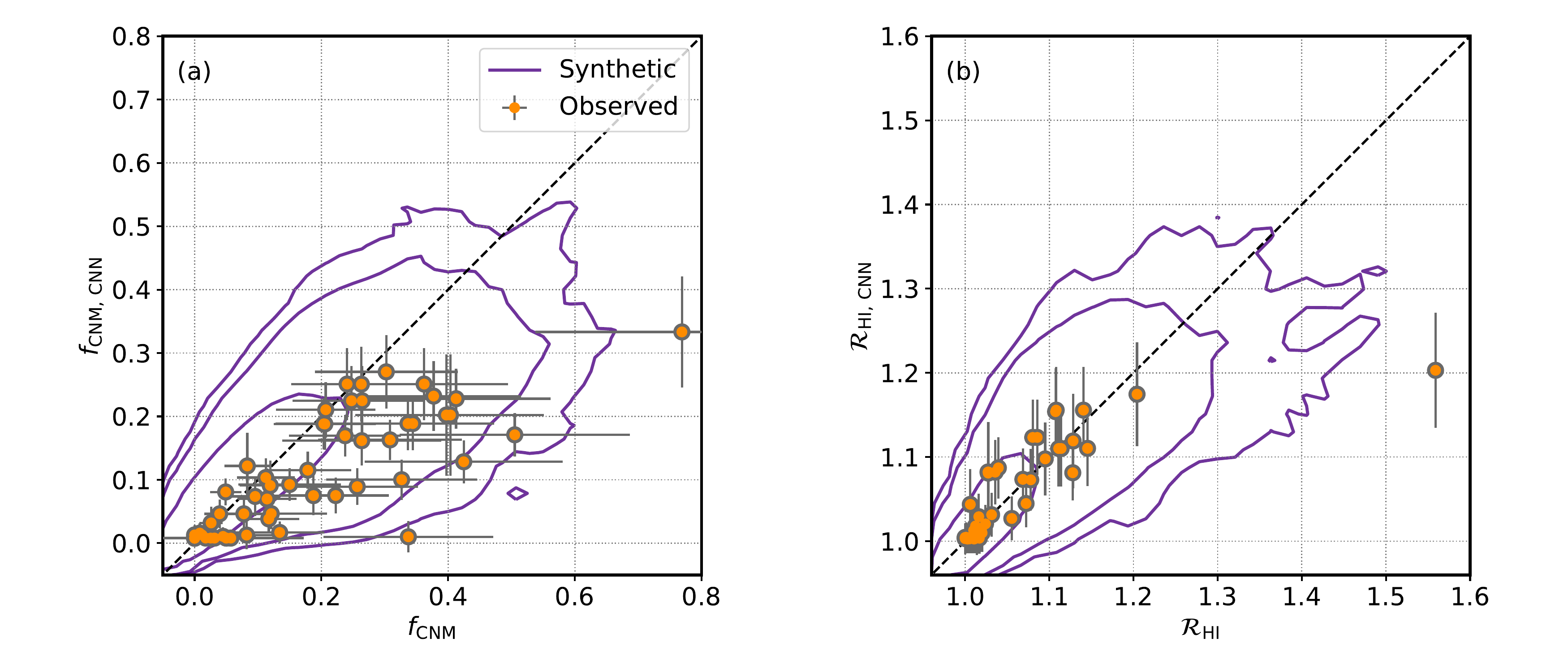}
\caption{Comparison of the ``input" values of \fcnm\ (a) and \rhi\ (b) for the $8173$ synthetic spectral line pairs comprising the validation set (purple contours) and the \nabs\ observed spectral line pairs comprising the test set (orange points) with the associated CNN model predictions (``CNN"). The contours in each panel denote  $68,\,95,$ and $99\%$ of the synthetic distributions.} 
\label{f:residuals}
\end{center}
\end{figure*}

\begin{equation}
    f_{\rm CNM} \simeq \frac{T_{\rm CNM}}{\langle T_s \rangle}
    \label{e:fcnm_old}
\end{equation}

\noindent where $T_{\rm CNM}$ is the CNM kinetic temperature and $\langle T_s \rangle$ is the optical depth weighted average spin temperature along the LOS, given by,


\begin{equation}
    \langle T_s \rangle = \frac{\int \tau_{\rm HI}\,T_s\,\,dv}{\int \tau_{\rm HI}\,\,dv}.
\end{equation}

Comparing the ``true" \fcnm\ along their simulated LOS with Equation~\ref{e:fcnm_old} from their synthetic $21\rm\,cm$ spectral pairs, KOK14 determined that, for \fcnm$<0.2$ (i.e., the majority of LOS under consideration in this work), including an additional term to incorporate the spin temperature of the WNM ($T_{s,w}$) provides a better \fcnm\ approximation, given by (their Equation 12),

\begin{equation}
    f_{\rm CNM} \approx \frac{T_c}{\langle T_s \rangle} \frac{T_{s,w} - \langle T_s \rangle}{T_{s,w}-T_c}.
    \label{e:fcnm}
\end{equation}

\noindent For our analysis, we use Equation~\ref{e:fcnm} to compute \fcnm. We set $T_{\rm CNM} = 50\rm\,K$, which has been observed as a suitable approximation for the local ISM \citep[e.g.,][]{stanimirovic2014,murray2018b}, and $T_{s,w}=1500\rm\,K$ (KOK14). We note that $T_{s,w}$ does not necessarily correspond to the real spin temperature of the WNM, which can be higher than $1500\rm\,K$ \citep{murray2015}, but a reference temperature above which $f_{\rm CNM}$ becomes zero. 

The uncertainties in \rhi\ and \fcnm\ are determined by a simple Monte Carlo simulation. In each of $10^5$ trials, we compute \rhi\ and \fcnm\ using Equation~\ref{e:rhi} and~\ref{e:fcnm} after adding $j*\sigma_{\tau, \rm HI}(v)$ and $j*\sigma_{T_B,\rm exp}(v)$ to \thi\ and \tbexp\ respectively, where $j$ is drawn randomly from a uniform distribution between $\pm3$. For \fcnm, to include uncertainty in our choices of $T_c$ and $T_{s,w}$ in Equation~\ref{e:fcnm}, in each trial we select value from within the range of realistic values \citep[$20<T_c<150\rm\,K$; e.g.,][and $1000<T_{s,w}<6000\rm\,K$]{dickey2000}. The final uncertainties are computed as the standard deviation over all trials. Table~\ref{tab:sources} (Appendix~\ref{ap:sources}) includes values for \nhithin, \nhi, \rhi, and \fcnm\ the \nabs\ observed LOS.

In Figure~\ref{f:properties}, we display cumulative distribution functions (CDFs) of \fcnm, \rhi, and $N_{\rm abs}\times \sin{|b|}$ for the synthetic and observed samples of $21\rm\,cm$ spectral line pairs. For the synthetic data, we include CDFs for the original synthetic sample (i.e., with only noise added, not shifted and/or flipped spectra) and the augmented synthetic sample. The majority of gas probed by the high-latitude observations features \fcnm$\lesssim0.3$ and \rhi$<1.2$. We find that observed and augmented synthetic distributions agree well within uncertainties (computed via bootstrapped re-sampling of the \nabs\ observed LOS), whereas the original synthetic sample features significantly fewer components per LOS.  

\section{Analysis: The Convolutional Neural Network}
\label{sec:CNN}

\begin{figure*}
\begin{center}
\includegraphics[width=\textwidth]{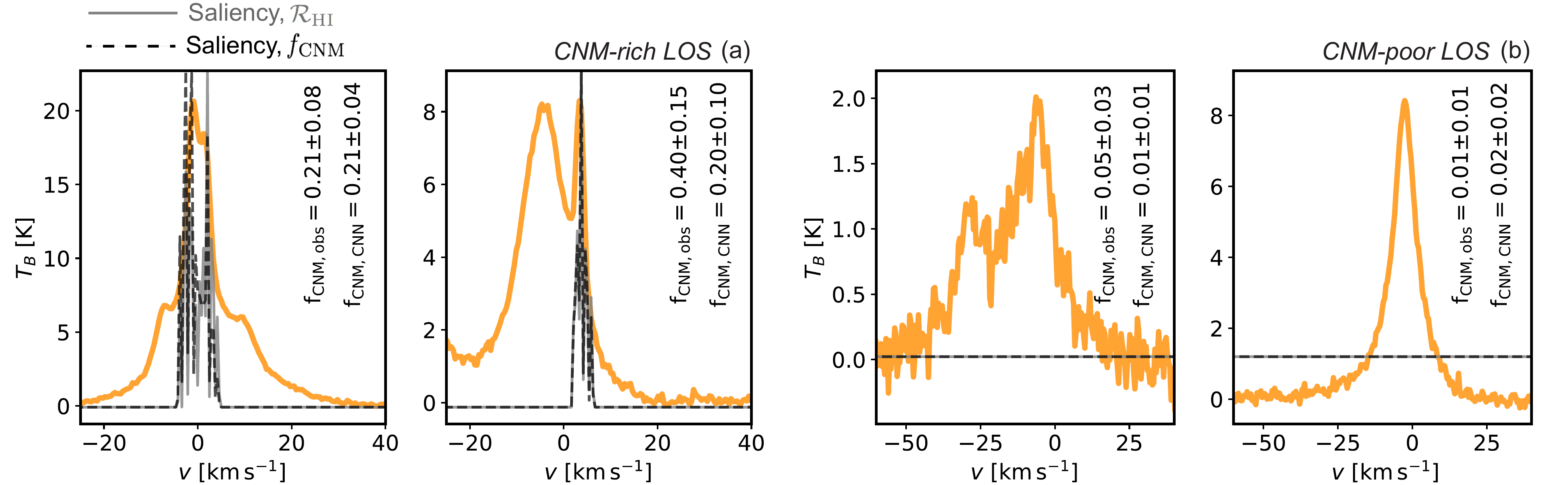}
\caption{Illustration of the salient components of the observed validation spectra (orange) which drive the predictions of \fcnm\ and \rhi. We compute the ``saliency" (see Section~\ref{sec:saliency}) of the final fully-connected output layer for \fcnm\ (grey) and \rhi\ (black dashed) for two examples of CNM-rich LOS (a), and two examples of CNM-poor LOS (b). For \fcnm, we include the observed ($f_{\rm CNM, abs}$) and predicted ($f_{\rm CNM,CNN}$) values and uncertainties within each panel. We find consistent results between the observed and synthetic samples. The saliency spectra are arbitrarily scaled.  }
\label{f:saliency}
\end{center}
\end{figure*}

To build a model which is able to accurately predict the thermodynamic state of \hi\ in the local ISM (parameterized here by \rhi\ and \fcnm), we use a deep neural network constructed using the Tensorflow and Keras frameworks \citep{chollet2015, tensorflow2015}. A neural network consists of layers of artificial ``neurons", each of which accepts input from neurons in the previous layer, and generates output information via an activation function \citep[for an early description in the astronomy literature, see][]{dieleman2015}. The activation function mimics the performance of biological neurons and is typically monotonically-increasing and non-linear. For example, a common activation function is linear rectification ($f(x) = \max(x, 0)$).  A neuron's output, $y$, is given by,

\begin{equation}
    y = f \left ( \sum x_i \cdot w_i + b \right )
\end{equation}

\noindent where $f$ is the activation function, $x_i$ are the inputs, $w_i$ are the weights associated with each input, and $b$ is a bias offset. A standard network consists of an input layer, multiple ``hidden" layers, and an output layer featuring the predicted values. In the case of a CNN, the network may contain convolutional layers which map topological structures between layers by convolving the inputs \citep{fukushima1980, lecun1998, dieleman2015}. A layer is ``fully-connected" if all neurons in the layer are connected to every neuron in the previous layer.

To determine the optimal weights and biases connecting the neurons in each layer, the network must be trained using a representative sample with known input and output values. After running the sample through the network (initialized with random values for the weights and biases), the output values are compared to the ``true" inputs via a loss function, and the weights and biases are adjusted via back-propagation of error with gradient descent.

For this work, following previous efforts to use CNNs to analyze stellar spectra \citep{li2017, wang2017, fabbro2018}, we construct a simple CNN using a combination of convolutional and fully-connected layers. The architecture of the CNN is illustrated in Figure~\ref{f:architecture}. The input layer ($1\times414$) consists of the $T_{B,\rm syn}(v)$ spectrum. The next two layers are convolutional layers with $8$ filters each and activation via linear rectification functions. These layers reduce the dimension of the input spectrum using convolutional windows of length $6$ and $40$ channels respectively. Next, we randomly remove $30\%$ of the output values of the second convolutional layer with a ``dropout" layer. This is to improve regularization and minimize over-fitting or ``co-adaptation" between neurons \citep{srivastava2014}. Finally, the output layer is fully-connected, with a linear activation function. 

\subsection{Training}

To train the CNN, we use the synthetic $21\rm\,cm$ dataset constructed in Section~\ref{sec:synthetic_spectra}. We start by randomly selecting $70\%$ of the sample for training, and reserve the remaining $30\%$ (\nval\ spectra) to test the CNN's performance. The training process involves feeding the sample of training data through the CNN: the weights and biases at each layer are initially set randomly, and the output is an estimate of \rhi\ and \fcnm\ for each input synthetic spectrum. This output is then compared to the ``true" input values using the mean-squared error (MSE) loss function, and the weights and biases are iteratively adjusted via back-propagation to minimize the MSE. Training concludes when a minimum value of the MSE is reached following $8$ epochs. When the network is applied to the synthetic test dataset (i.e., the $30\%$ reserved and not used for training), we achieve $\rm MSE=0.007$.

After training and testing the network with the synthetic dataset, we use the \nabs\ observed spectral line pairs as validation. We apply the trained network to the observed validation spectra, and find $\rm MSE=0.010$, which is consistent with the results of applying the network to the synthetic test dataset. 

\subsection{Estimating uncertainty}
\label{sec:uncertainty}

The uncertainty in the CNN predictions includes contributions from the trained weights connecting the network layers, the CNN architecture (e.g,. size and shape of the layers, choice of activation and cost functions) and the fidelity of the training dataset. To quantify the uncertainty due to the trained parameters, whose random initialization affects which local maximum in likelihood space the training will converge on, we re-train the CNN 
over a series of $25$ trials and store the trained weights and biases. We find consistent training performance (measured via MSE) across these trials. The final reported values and uncertainties of \rhi\ and \fcnm\ quoted in our results are the mean and standard deviation across these trials. We emphasize that these uncertainties are lower limits, as they do not encompass the biases and uncertainties inherent in the network architecture.

To assess the uncertainty due to the network architecture, we perform ten-fold cross-validation. Specifically, we split the full training dataset into ten groups, and iteratively re-train the network with nine groups and test it on the tenth group until all groups are tested. The mean absolute error in the network predictions following these iterations is $4.38\pm0.29\%$ (mean and standard deviation).

\begin{figure*}
\begin{center}
\includegraphics[width=\textwidth]{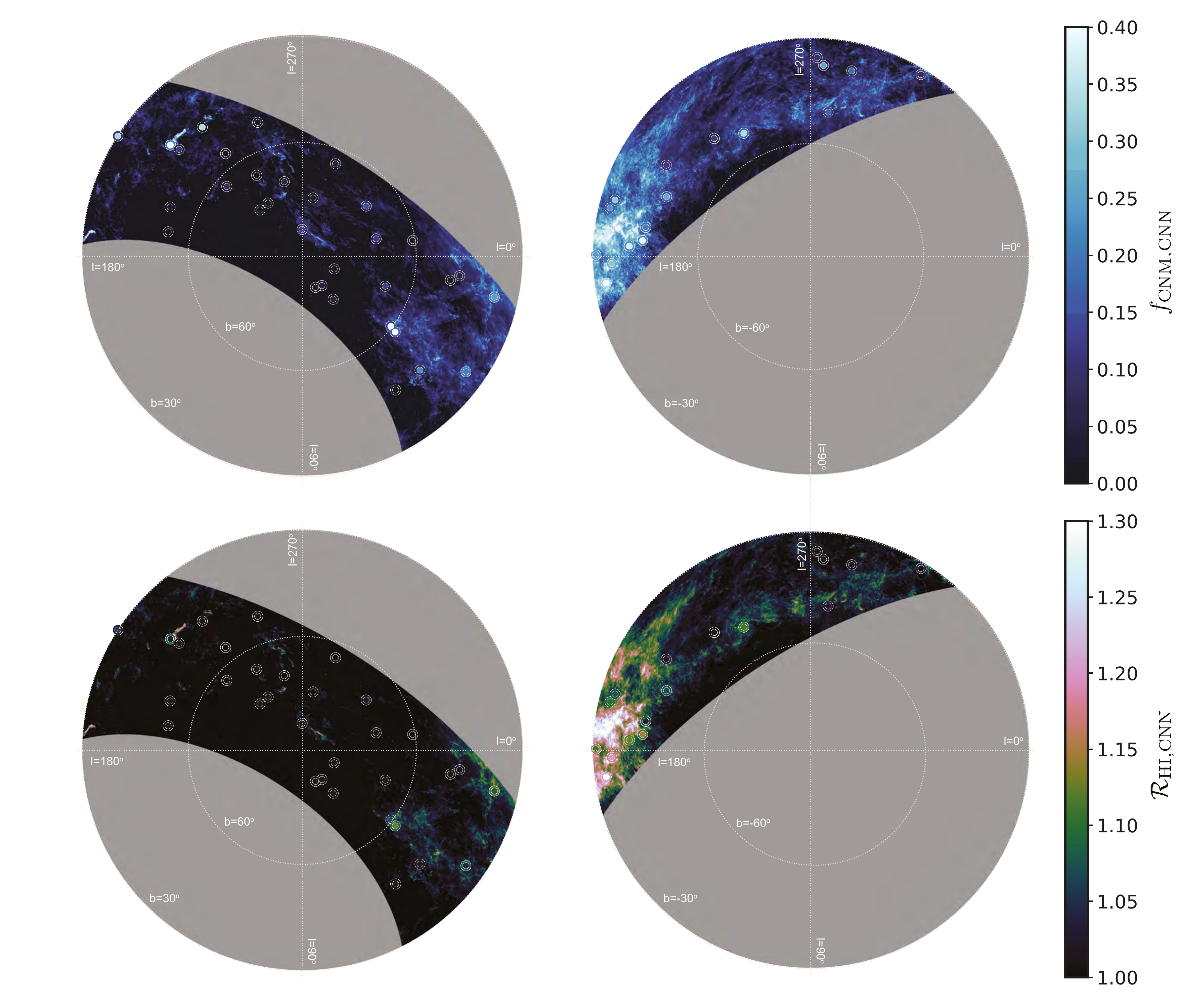}
\caption{Maps of \fcnm\ (top) and \rhi\ (bottom) predicted by the trained CNN for the full high latitude sky ($|b|>30^{\circ}$; NGP left, SGP right) observed by GALFA-\hi, computed as the median values following the 25 training trials. The observed LOS from the validation set are included as circles, which are colored by the observed values (inside circles) and the CNN prediction (outside circles).}
\label{f:cnn_maps}
\end{center}
\end{figure*}

All of the training and test data, as well as the code used to build, train and evaluate the CNN, and generate resulting figures are publicly available.\footnote{Training data: \url{https://doi.org/10.7910/DVN/QT6NPF}; Observed catalog: \url{https://doi.org/10.7910/DVN/MJGQAY}; Software: \url{https://doi.org/10.5281/zenodo.3923100}}

\section{Results}
\label{sec:results}

In Figure~\ref{f:residuals}, we compare the predicted values of \rhi\ and \fcnm\ from the trained CNN (``output") with the input values for the synthetic test set (\nval\ spectra) and the \nabs\ observed spectral line pairs. We observe that the CNN does a reasonable job of reproducing the observed and simulated \fcnm\ and \rhi\ for the majority of each sample, within uncertainties. At low \fcnm\ and \rhi, although the observed (input) and predicted CNN results agree within uncertainties, the CNN converges most data around values of \fcnm$\sim0$ and \rhi$=1.0$. At high \rhi\ and \fcnm, the CNN tends to under-predict the input values. This is likely due to the fact that, despite augmentation, the spectra used for training have more low-\fcnm\ LOS (c.f., Figure~\ref{f:properties}). 

\begin{figure*}
\begin{center}
\includegraphics[width=\textwidth]{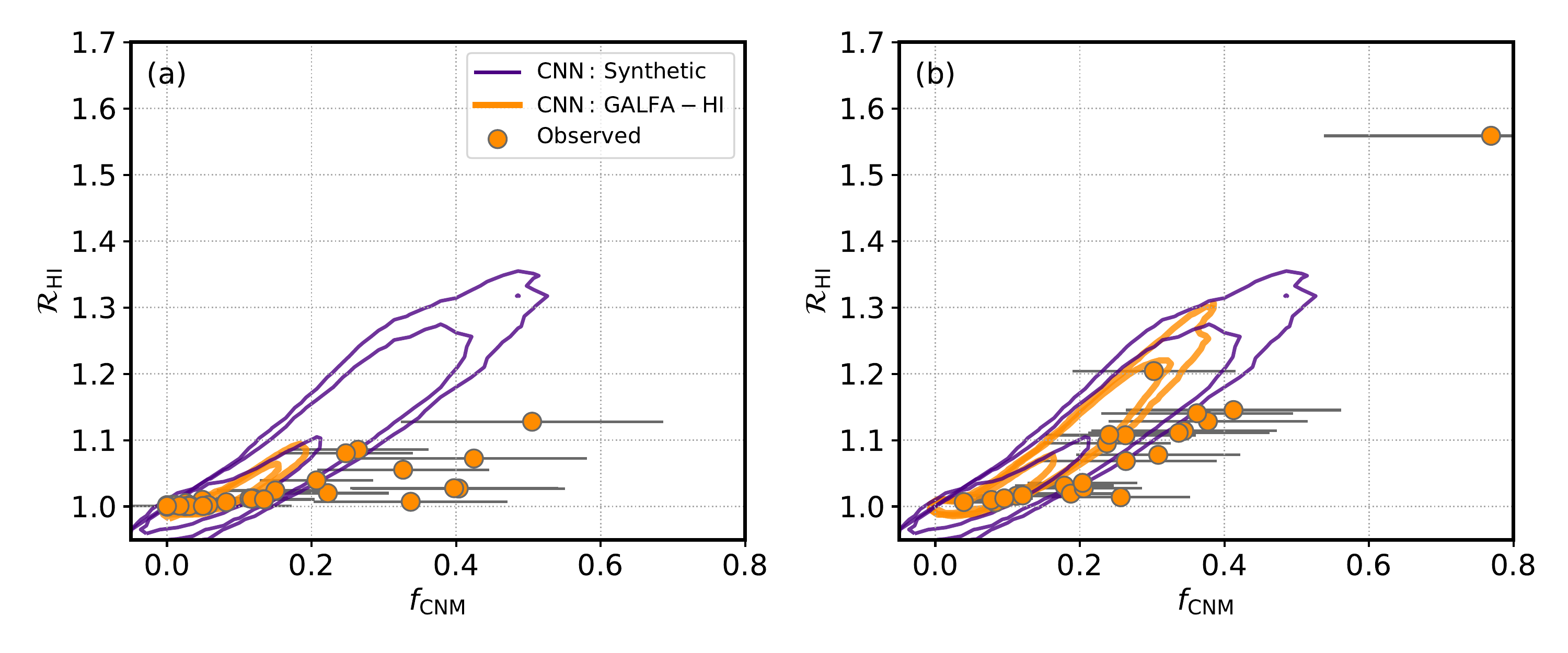}
\caption{\rhi\ versus \fcnm\ for the CNN predictions for GALFA-\hi\ (orange contours) for the Galactic Northern (a) and Galactic Southern (b) hemispheres. We compare these distributions with observed values of \fcnm\ and \rhi\ from the \nabs\ validation targets (orange points). As expected, LOS with more CNM correspond to larger column density correction factors due to \hi\ optical depth. We include the distribution of \rhi\ vs. \fcnm\ predicted by the CNN for the synthetic test dataset (purple contours). Contours in each panel indicate the $68$, $95$ and $99\%$ limits of the CNN distribution. The innermost contour for the GALFA-\hi\ distribution in the Galactic Northern hemisphere (a) is not visible from behind the data points in this figure, as $\sim 68\%$ the data are predicted to have \rhi$=1$ and \fcnm$=0$.} 
\label{f:rf_comparison}
\end{center}
\end{figure*}

\subsection{Saliency}
\label{sec:saliency}

A common issue with interpreting the results of CNNs is the difficulty in identifying what the CNN is actually learning from the input training data. To gain intuition about which features of the data are most important for driving the output predictions, we use a method called ``saliency" \citep{simonyan2013}. The saliency is computed as the derivative of the output prediction (i.e., \fcnm\ or \rhi) with respect to the spectral channel values for an individual layer in the CNN. This quantifies which channels most significantly alter the output prediction from the layer if changed. Saliency spectra \citep[or maps, in the case of 2D neural networks; e.g.,][]{peek2019a} are therefore useful not only for gaining physical intuition about relevant spectral features, but also for diagnosing problems with the CNN, which may end up inadvertently prioritizing spurious artifacts for generating predictions.

To further investigate which features of the data motivate the results in Figure~\ref{f:residuals}, we compute the saliency of the final fully-connected layer in the CNN (i.e., the output prediction layer) for the observed validation data, using the Keras framework. We display examples of the results in Figure~\ref{f:saliency}. We select two representative CNM-rich and CNM-poor LOS from the observed sample. In each case we compare the input spectrum with the saliency spectrum for \fcnm\ and \rhi. For CNM-rich LOS (defined here as \fcnm$\gtrsim0.2$) the CNN is sensitive not only to the peak, but also to the shape of each component, including their edges and widths, in agreement with theoretical expectations. For the CNM, we expect thermal broadening to be limited by low kinetic temperatures, and therefore we expect narrow ($\delta v \sim 1-5\rm\,km\,s^{-1}$) spectral line widths. Accordingly, we observe from the saliency spectra that the CNN is particularly sensitive to the presence of narrow components when predicting \fcnm\ and \rhi. However, when the profile is complex, the CNN can over-interpret sharp spectral edges. For example, as shown in Figure~\ref{f:saliency}a, when a CNM-rich LOS includes high-amplitude, sharp, narrow features, the CNN prioritizes the edges of these features over the presence of additional, significant lower-amplitude emission, resulting in an under-prediction of \fcnm.  In general, from the saliency spectra it is clear that the network tends to prioritize channels with $|v|\lesssim 20\rm\,km\,s^{-1}$ where $T_B$ peaks. Improving the augmentation of the synthetic spectra to account for the presence and influence of emission at higher velocities (i.e., $|v|>20\rm\,km\,s^{-1}$) will be required in future work. We display saliency results for all \nabs\ LOS in Figure~\ref{f:sal_maps_all} (Appendix~\ref{ap:sal}).

In the case of CNM-poor LOS, the CNN is equally sensitive to all channels, or to all channels with significantly-detected emission. We find consistent results for saliency spectra computed for the synthetic spectra. 

\begin{figure*}
\begin{center}
\includegraphics[width=\textwidth]{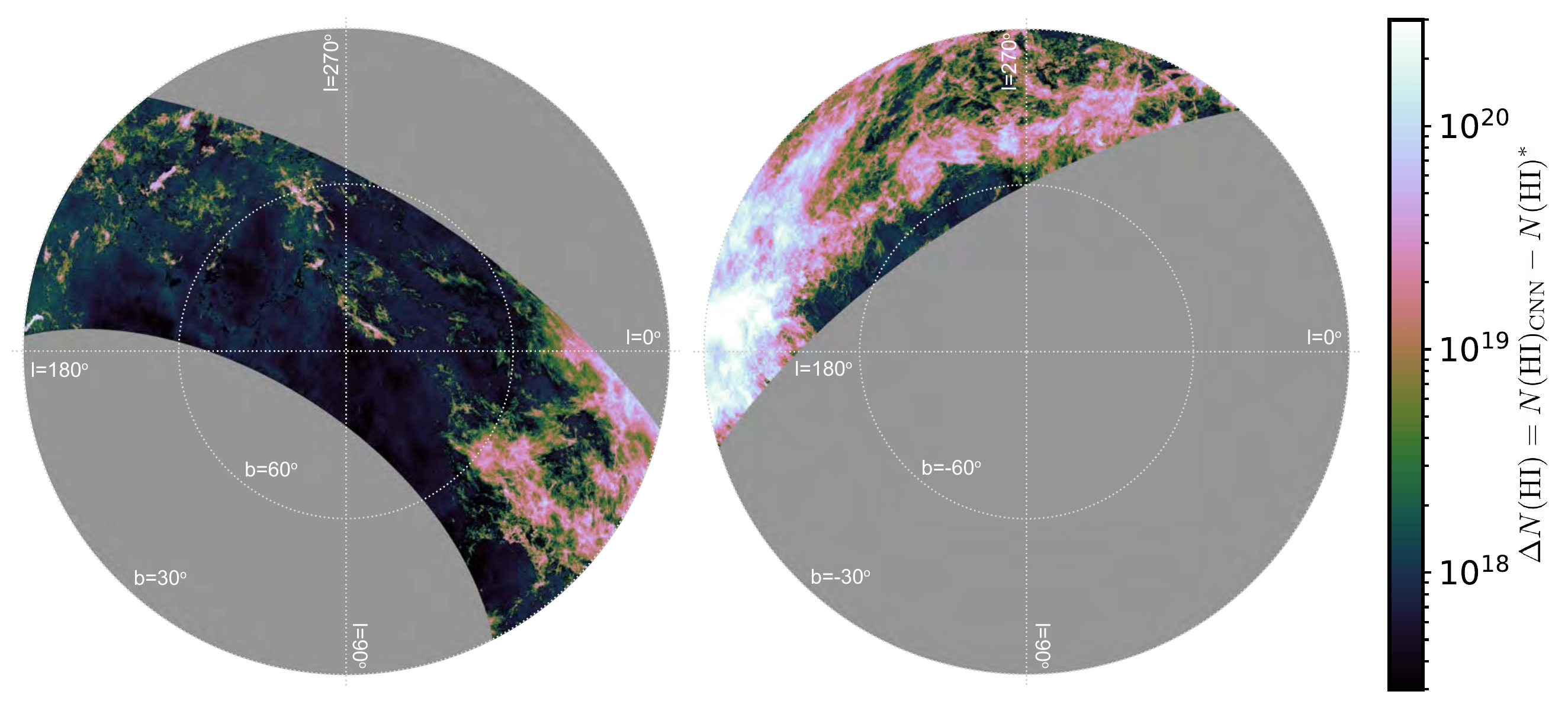}
\caption{Maps of $\Delta N({\rm HI})$: the difference between \nhi\ and \nhithin, tracing the \hi\ column density ``missed" in the optically-thin limit as a result of optically-thick \hi. }
\label{f:nhi_resids}
\end{center}
\end{figure*}

\subsection{CNM maps from the CNN}

We next apply the CNN to the full GALFA-\hi\ survey to generate all-Arecibo-sky (at $|b|>30^{\circ}$)  maps of \fcnm\ and \rhi\ and their uncertainties predicted by the CNN. These maps are displayed in Figure~\ref{f:cnn_maps} and Figure~\ref{f:cnn_uncertainties} (Appendix~\ref{ap:uncert_maps}). We include the \nabs\ observed LOS in these maps, where the interior of each target is colored by the observed \fcnm\ and \rhi\ values, and the exterior is colored by the predicted \fcnm\ and \rhi\ values. The predicted values are computed as the average value within a $9\times9$ pixel grid around each source, excluding the central $3\times3$ pixel region (i.e., consistent with the method of extracting \tbexp\ for each LOS from GALFA-\hi; Section~\ref{sec:sample}).

By comparing Figure~\ref{f:cnn_maps} and Figure~\ref{f:cnn_uncertainties} by eye, we observe that the CNN uncertainties increase with increasing \fcnm\ and \rhi. This is due to the fact that in the presence of cold \hi, the observed (and synthetic) emission spectra exhibit stronger complexity in the form of blended spectral features and more diverse feature shapes, resulting in increased uncertainties in the CNN predictions. In addition, the synthetic dataset is dominated by low-\fcnm\ LOS (Figure~\ref{f:properties}), which results in increased precision at low \fcnm.  We emphasize that, given inherent uncertainties in estimating low-\fcnm\ ($\lesssim0.1$) from line pairs using Equation~\ref{e:fcnm} (KOK14) and high \fcnm\ from the CNN due to spectral complexity, that we are most confident in predicting intermediate values of \fcnm\ ($\sim0.1\lesssim$\fcnm$\lesssim0.2$) in this work. We discuss these effects further in Appendix~\ref{ap:uncert_maps}.  

The large-area maps of \fcnm\ and \rhi\ in Figure~\ref{f:cnn_maps} indicate that cold neutral gas structures are found ubiquitously, even in the high-latitude sky ($|b|>30^{\circ}$). We successfully recover known cold \hi\ structures, and resolve complex structures connecting them on large angular scales. These maps and their uncertainties are publicly available.\footnote{Maps: \url{https://doi.org/10.7910/DVN/E0HLON}}

In Figure~\ref{f:rf_comparison}, we plot the relationship between \fcnm\ and \rhi\ for the CNN applied to GALFA-\hi, and compare the distributions with the ``true" input values for the \nabs\ observed LOS. We split the results into two panels to compare the Galactic Northern and Southern hemispheres. We observe that, as expected, as \fcnm\ increases, so does \rhi\ (i.e., more cold, optically-thick \hi\ corresponds to larger correction factors for optical depth to \nhi). The model results also follow the same trends, within uncertainties, as the observed sample, lending further confidence that the training set provides a reasonable model of observed \hi\ properties. We include the distribution of predicted \rhi\ vs. \fcnm\ by the CNN for the synthetic test set in each panel of Figure~\ref{f:rf_comparison}. The CNN predictions for GALFA-\hi\ appear as a subset of the predictions for synthetic distribution. 

\begin{figure*}
\begin{center}
\includegraphics[width=\textwidth]{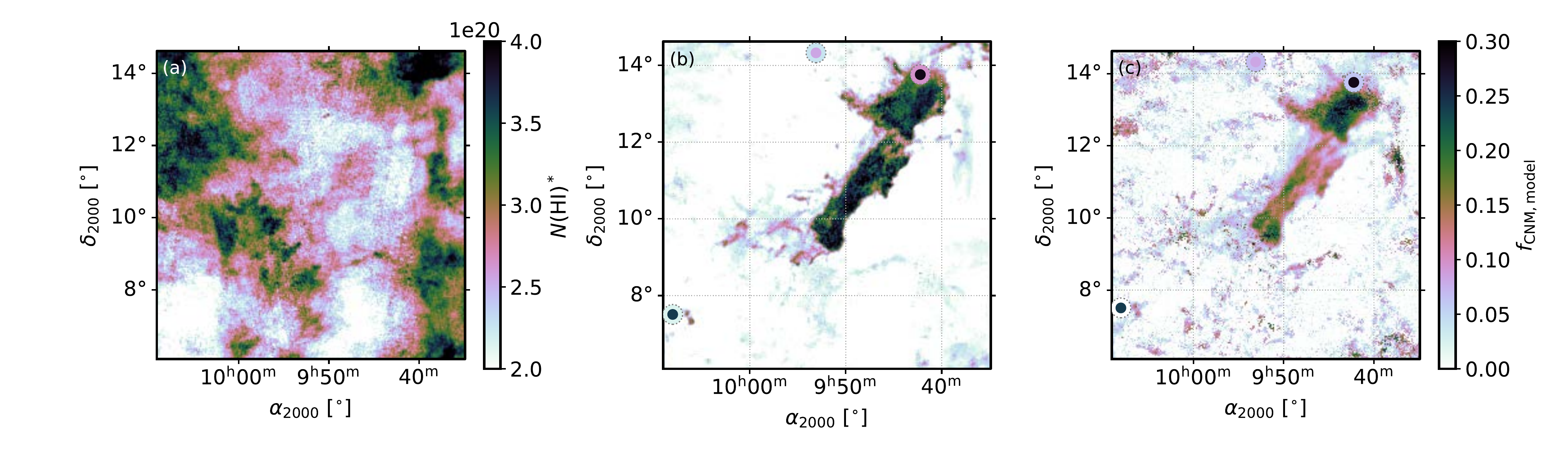}
\caption{A test region to assess the performance of the CNN. (a) The \hi\ column density (in the optically-thin limit, \nhithin; (b) $f_{\rm CNM}$ from the CNN; (c) $f_{\rm CNM}$ from the GaussPy+ decomposition (Appendix~\ref{ap:gp}). The CNN and decomposition methods both successfully extract the Local Leo Cold Cloud \citep[LLCC; e.g.,][]{peek2011}, the coldest-known \citep[$T\sim20\rm,K$][]{heiles2003} in the local ISM. The observed LOS from the validation set are included as circles which are colored by the observed \fcnm\ (inside circles) and the CNN (panel b) and GaussPy+ (panel c) results (outside circles).}
\label{f:llcc}
\end{center}
\end{figure*}

In addition, we observe from Figure~\ref{f:rf_comparison}, in agreement with a by-eye analysis of  Figure~\ref{f:cnn_maps}, that the Southern high-latitude Galactic hemisphere observed by GALFA-\hi\ contains significantly more CNM. The Arecibo Observatory declination range in the Southern hemisphere includes the Taurus-Perseus molecular cloud complex (although the main bodies of both clouds are excluded by our $|b|>30^{\circ}$ latitude cut) as well as the Orion-Eridanus super bubble wall, which is known to contain cold, dense gas \citep{heiles1999}. 

To illustrate the amount of \hi\ ``missed" by the optically thin approximation in this environment, in Figure~\ref{f:nhi_resids} we map the residual \hi\ column density after applying the optical depth correction via \rhi. This residual is computed as the difference between the corrected and uncorrected \hi\ column density. Figure~\ref{f:cnn_maps} indicates that the correction factor for optically-thick \hi\ is generally small, and Figure~\ref{f:nhi_resids} illustrates that across large areas of sky (particularly in the north) the missing column density is $\sim10^{18}\rm\,cm^{-2}$. However, in regions of sky featuring well-known cold structures, the ``missing" column can reach $>10^{20}\rm\,cm^{-2}$. However, the cumulative mass fraction of this ``missing" \hi\ for the high-latitude GALFA-\hi\ sky is $\sim2\%$ and $\sim8\%$ for the North and South respectively, or $\sim5\%$ overall.

\subsection{Comparison with Gaussian Decomposition}

A common method for quantifying \hi\ properties from $21\rm\,cm$ emission observations is to decompose the spectral features into Gaussian functions and then estimate the temperature and column density of each feature using its spectral line width and amplitude \citep[e.g.,][]{takakubo1966, mebold1972, haud2007,kalberla2018, marchal2019}. As mentioned previously, this method typically requires subjective input for selecting component parameters. However, as CNNs have, to our knowledge, never been applied in this context, it is of interest to compare our results with the ``state of the art" Gaussian decomposition method. 

To conduct the comparison, we decompose GALFA-\hi\ data using GaussPy+ \citep{riener2019}\footnote{https://github.com/mriener/gausspyplus}, a Python package based on the Autonomous Gaussian Decomposition \citep[AGD;][]{lindner2015} algorithm. In Appendix~\ref{ap:gp} we describe the decomposition process, and the method of estimating \fcnm\ from the results. We find that the CNN performs equivalently well, even slightly better than the Gaussian decomposition method at predicting \fcnm\ constrained by $21\rm\,cm$ absorption observations.

As another test, we zoom-in on the Local Leo Cold Cloud \citep[LLCC;][]{verschuur1969, meyer2006,peek2011}. Located at a distance of $\sim20\rm\,pc$, the LLCC is the coldest-known \hi\ cloud in the local ISM, with a temperature of $\sim20\rm\,K$ \citep{heiles2003}. In Figure~\ref{f:llcc} we compare the total \hi\ column density map (a) with the \fcnm\ maps from the CNN (b) and the GaussPy+ decomposition (c) of the LLCC. The CNN map agrees well with the GaussPy+ map, as well as with previous models of the cloud using by-hand Gaussian decomposition \citep{peek2011}. In addition, the CNN appears to more accurately predict \fcnm\ for the CNM-rich sightlines intersecting the LLCC. However, both methods appear to underestimate the \fcnm\ inferred from $21\rm\,cm$ absorption. This discrepancy is likely driven by the relative lack of CNM-rich LOS and lack of features at high velocities (e.g., $|v|\gtrsim20\rm\,km\,s^{-1}$ in the synthetic training set, which makes this model insufficient for modeling similarly complex spectral features. Future CNNs trained by next-generation simulations will address this bias directly.

Overall, we find at least consistent (if not slightly better) agreement between \fcnm\ from $21\rm\,cm$ absorption measurements with the CNN as with Gaussian decomposition. Further improvements can be made to both methods by including information about the morphology of \hi\ structures, which is known to contain significant information about their physical properties \citep{clark2014, clark2019, peek2019a, peek2019b}.

\subsection{The nature of small-scale HI structures}

\begin{figure}
\begin{center}
\includegraphics[width=0.4\textwidth]{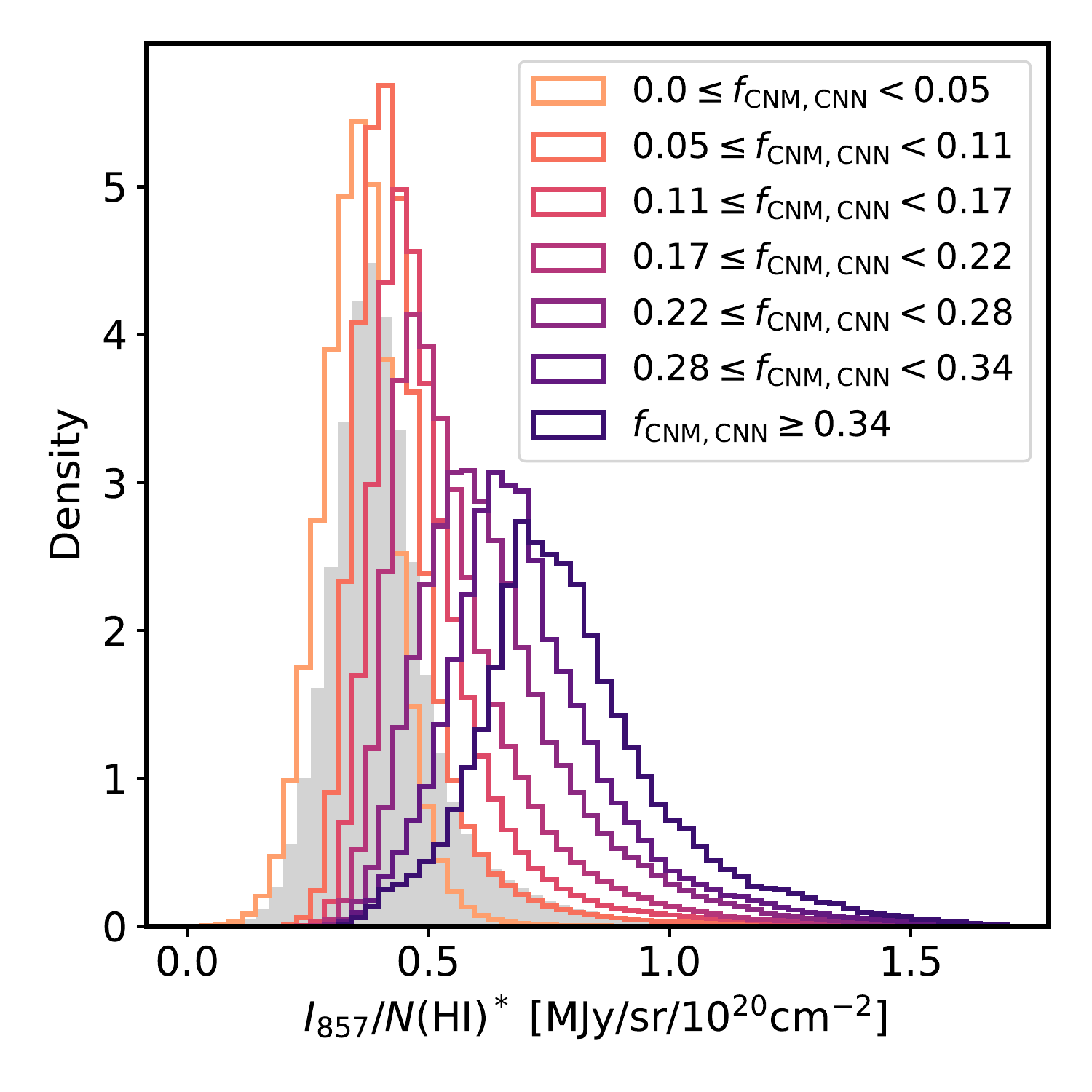}
\caption{Histogram of $I_{857}/$\nhithin\ for the $|b|>30^{\circ}$ sky observed by GALFA-\hi\ (gray). We divide the sky into equally-spaced bins of $f_{\rm CNM}$ up to $f_{\rm CNM}=0.4$ and plot histograms of $I_{857}/$\nhithin\ as a function of $f_{\rm CNM}$ (colored according to inset legend).}
\label{f:fir_ratio}
\end{center}
\end{figure}

The high-latitude, large-area maps of \fcnm\ and \rhi\ constructed by the CNN are useful not only for illustrating the spatial distribution of CNM-rich structures, but also for considering statistical properties of local Galactic \hi. In particular, we can test hypotheses related to the physical properties of distinct \hi\ morphological phases.

High-resolution observations have established that \hi\ in the local ISM exhibits remarkably linear, filamentary structures which are aligned with the magnetic field in the plane of the sky traced by starlight and dust polarization \citep{clark2014, clark2015}. The origin of these features is contentious -- either they are caused by real density structures, or they arise due to velocity caustics created by the turbulent velocity field in the ISM \citep[e.g.,][]{lp2000}.

\begin{figure*}
\begin{center}
\includegraphics[width=0.98\textwidth]{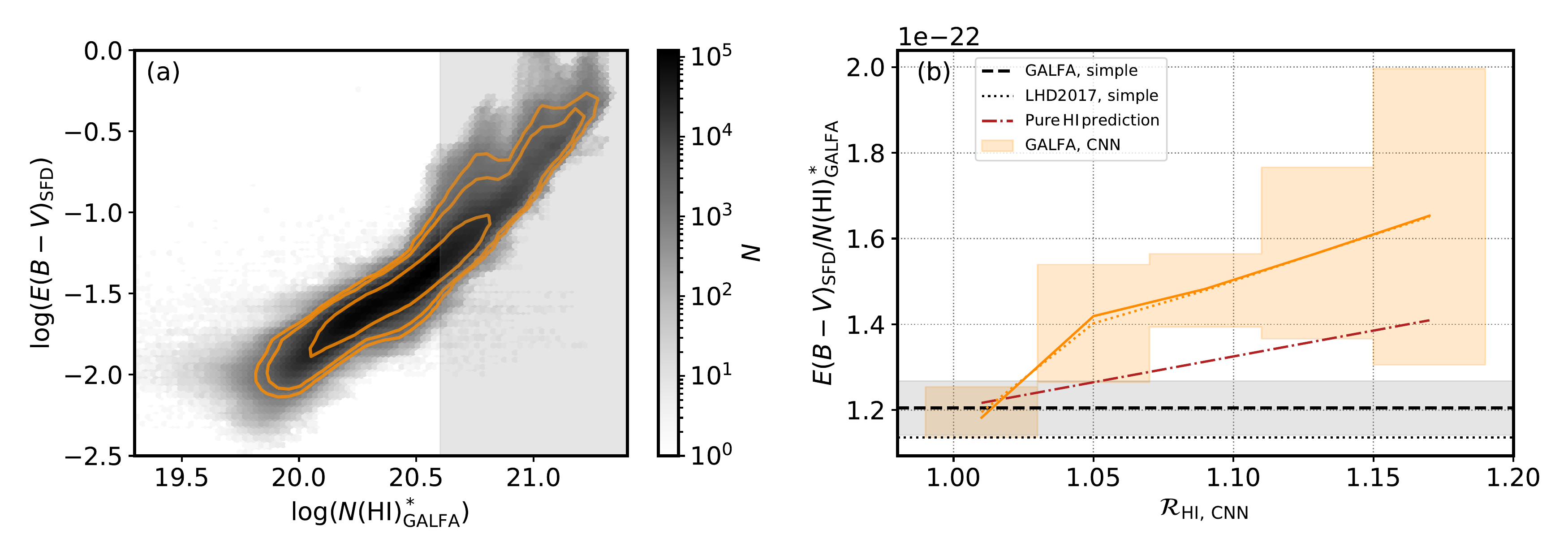}
\caption{Investigating the relationship between \hi\ column density and dust reddening at high $b$. (a): \nhithin\ from GALFA-\hi\ vs. \ebv\ \citep{sfd98}. We highlight the low-column density regime identified by LHD17 as being atomic gas-dominated (i.e., higher \nhithin\ is shaded). Contours illustrate the $68$, $95$ and $99\%$ limits of the observed distribution (b): The \nhithin/\ebv\ slope as a function of \rhi. We include the simple approach of fitting all gas with \nhithin$<4\times10^{20}\rm\,cm^{-2}$ (black dashed) and the result of LHD17 (black dotted), observing that although the estimates differ, having been computed using different data sets (GALFA-\hi\ vs. HI4PI), they are consistent within the block-bootstrapped uncertainties of the GALFA-\hi\ value. We compare these values with the result of fitting slopes within bins of increasing \rhi\ (orange) and the pure \hi\ prediction of modifying the simple result by the median \rhi\ per bin. The uncertainties for the simple and CNN results are computed by block-bootstrapping the GALFA-\hi\ sky (see text). In addition, we repeat the analysis after masking all pixels with significant CO emission and find no significant change (orange, dotted). }
\label{f:ebv_ratio}
\end{center}
\end{figure*}

Recently, \citet{clark2019} resolved this debate by establishing that small-scale, linear features observed in \hi\ $21\rm\,cm$ channel maps must be real density structures. They showed that small-scale \hi\ structures are correlated with far infrared (FIR) emission intensity, which traces the density field rather than the velocity field, and that the FIR/\nhi\ ratio associated with these features increases with increasingly small scales. In agreement, \citet{kalberla2020} find consistent variation of FIR/\nhi\ with small scale structure intensity. These findings support the picture that filamentary small-scale structures originate preferentially in real, cold, dense neutral gas (i.e., CNM). In addition, \citet{peek2019b} showed that the equivalent width of Na I absorption (a tracer of the cool ISM) depends more strongly on the prevalence of small-scale structure than on the total column density, which yet further supports the picture that small-scale \hi\ structures are dominated by the CNM. 

The \fcnm\ maps presented here provide another test of this result. Specifically, we can test how the ratio of FIR emission and \hi\ column density (tracing the density field) vary with increasing \fcnm, similar to \citet{clark2019}. To trace FIR emission, we use the all-sky map at $857\rm\,GHz$ from Planck\footnote{Data release R3.01} \citep{planck2018}. Following \citet{clark2019}, we subtract a monopole correction of $0.64\,\rm MJy/sr$ \citep{planck2016} from the map before re-projecting it into the high-latitude ($|b|>30^{\circ}$) GALFA-\hi\ footprint. In Figure~\ref{f:fir_ratio} we display histograms of $I_{857}$/\nhithin\ in bins of \fcnm. We observe that $I_{857}$/\nhithin\ increases with increasing \fcnm. Taken together with the results of \citet{clark2019}, who showed that $I_{857}$/\nhithin\ also increases with small-scale structure intensity (c.f., their Figure 8), this result is fully consistent with the picture that observed small-scale structures are preferentially CNM. 

\subsection{Calibrating the HI to reddening ratio}

In addition to diagnosing the nature of small-scale \hi\ structure, we can calibrate the use of \hi\ as a tracer of dust properties. Recently, \citet[][hereafter LHD17]{lenz2017} leveraged a simple linear relationship between \nhithin\ and \ebv\ at high Galactic latitudes (\nhithin/\ebv$=8.8\times10^{21}$) to produce a map of \ebv\ using data from the HI4PI survey with the highest fidelity at high latitude to date. This method assumes \hi\ traces the total gas column density (i.e., no contribution from molecular or ionized gas), \nhi=\nhithin, and also that a single relation applies everywhere. However, as we have shown using the \rhi\ map presented in Figure~\ref{f:cnn_maps}, although the overall correction is small, \rhi\ does vary across the high-latitude sky and therefore, generally, \nhi$>$\nhithin. In addition, dust properties vary between interstellar gas phases (e.g., as shown in Figure~\ref{f:fir_ratio}), and so it is not clear that a single linear relationship is warranted, even in the low-column density regime. 

To investigate further, we compare \hi\ column density and dust reddening (\ebv) as a function of the CNN-derived \hi\ properties. To trace \ebv, we use the map produced by \citet[][hereafter SFD98]{sfd98} using dust emission at $100\,\mu m$ from the Infrared Astronomy Satellite (IRAS) mission, which we re-project into the GALFA-\hi\ footprint. We isolate the low-\nhithin\ regime where LHD17 assume a simple linear relation between \nhithin\ and \ebv\ (i.e., \nhithin$<4\times10^{20}\rm\,cm^{-2}$). In Figure~\ref{f:ebv_ratio}(b) we plot the evolution of the \ebv/\nhithin\ slope within this range within bins of increasing \rhi. For each bin, we estimate the uncertainties on the slope by block-bootstrapping the GALFA-\hi\ sky using twelve blocks: six equally spaced from $0<\alpha_{2000}<360$ by two equally spaced from $1<\delta_{2000}<38^{\circ}$.\footnote{We tested different block configurations and found that the choice of blocks does not significantly affect our results.} We observe that there is a significant change in slope between the optically thin (\rhi=1.0) and the \rhi$ > 1$ regimes. This behavior validates the overall sense of our results, indicating that the CNN identifies cold neutral gas successfully. 

If the increase in \ebv/\nhithin\ with increasing \rhi\ is due to \hi\ optical depth effects alone, then the observed trend in Figure~\ref{f:ebv_ratio} should agree with the empirical prediction of modifying the simple linear slope by \rhi. Given that \nhi = \nhithin \rhi, multiplying the \ebv/\nhithin slope by \rhi\ within each bin should trace the true evolution of \ebv/\nhi\ (assuming the high-latitude sky is purely \hi).  We include this ``pure \hi" prediction in Figure~\ref{f:ebv_ratio}, and find that the observed increase in slope with \rhi\ exceeds the pure \hi\ prediction.

Any ISM phase beyond \hi\ in our region of interest will bias \ebv/\nhithin\ to spuriously high values (i.e., there would be more gas traced by \ebv\ than accounted for by \hi\ alone), and may explain the excess increase in \ebv/\nhithin\ shown in Figure~\ref{f:ebv_ratio}.
To test for the presence of molecular gas, we use the all-sky map of molecular gas emission in the form of $^{12}\rm CO\,J=1-0$ emission from \emph{Planck}
Commander foreground separation \citep{planck2015}. We reproject the HEALPix \citep{gorski2005} map with $N_{\rm side}=256$ to the GALFA-\hi\ footprint, and mask all pixels without significant detection (defined as $I_{\rm mean}/I_{\rm RMS}<3$ where $I_{\rm mean}$ and $I_{\rm RMS}$ are the mean and standard deviation of the posterior from the Commander analysis).
Although the agreement improves when we repeat our analysis after masking these regions, the change is not significant (c.f., Figure~\ref{f:ebv_ratio}). 

To explain the remaining discrepancy, it is likely that additional, complex effects are contributing. Detailed comparisons between sensitive tracers of the the total ISM column density at high latitude \citep[e.g., gamma rays, dust;][]{grenier2005,planck2016} and available gas tracers in emission (e.g., \hi, CO) reveal a significant population of gas without detectable neutral or molecular gas emission. Tracking down the origins of this ``dark" gas is the subject of considerable observational \citep[e.g.,][]{liszt2010,lee2015, remy2018} and theoretical \citep{glover2016, seifried2020} efforts. Available $21\rm\,cm$ absorption measurements at high Galactic latitudes have already ruled out the hypothesis that ``dark" gas can be accounted for by optically-thick \hi\ alone \citep{murray2018a}. The remaining possibilities include molecular hydrogen (H$_2$) undetected by CO emission, and variations in dust emissivity which complicate measurements of the total available column density. Although we cannot distinguish between these effects, our results support the picture that they are increasingly relevant in regions with CNM-friendly conditions (i.e.,  high-\fcnm\ and high-\rhi).

Overall, it is clear that a single linear relationship between \nhithin\ and \ebv\ is an oversimplification. Future efforts to predict \ebv\ with high fidelity using $21\rm\,cm$ surveys, even within high-latitude regions of interest for cosmological surveys, must take the effects of \hi\ phase structure into account. 

\section{Discussion}
\label{sec:discussion}

The success of the simple CNN at predicting \fcnm\ and \rhi\ indicates that, at the high Galactic latitudes studied here ($|b|>30^{\circ}$), there is information contained in the velocity structure of $21\rm\,cm$ emission spectra which reflect the absorption properties of the underlying medium. Furthermore, the agreement between the predicted distributions of \fcnm\ and \rhi\ from the synthetic and real observations indicates that the synthetic spectra feature realistic structure. The resulting large-area maps of \fcnm\ and \rhi\ are useful for diagnosing the ubiquity of CNM-rich structures in the local ISM, as well as statistically diagnosing the effect of \hi\ phase conditions on the relationship between \hi\ and dust properties at high latitude. 

Ultimately, this effort represents a proof of the concept that deep learning approaches can be used to quantify important ISM properties from spectral line data. There are clear areas in which the CNN model can be improved, including new training data from numerical simulations featuring increasingly realistic physical models in a wider range of Galactic environments (e.g., lower Galactic latitudes), and more sophisticated machine learning approaches such as generative networks for augmenting training data, higher-dimensional hybrid networks for including information about the morphology of \hi\ structures, and additional methods for interpreting the output of the trained network. However, that our simple CNN performs at least as well, if not better than current methods (e.g., diagnosing \hi\ conditions from decomposition of $21\rm\,cm$ emission spectra) already points to the potential for future studies, wherein multiple methods may be compared to diagnose biases and decipher realistic ISM properties. In addition, incoming surveys for $21\rm\,cm$ absorption \citep[e.g.,][]{dickey2013} will provide significantly more validation for future models. 

\section{Summary and Conclusions}
\label{sec:summary}

In this work, we show that a simple 1D CNN trained using synthetic $21\rm\,cm$ observations of numerical simulations (KOK14) can successfully predict \hi\ properties which formally require knowledge of the optical depth along the line of sight from $21\rm\,cm$ emission alone (Figure~\ref{f:residuals}), including the fraction of CNM along the line of sight (\fcnm) and the correction for optically-thick gas to the total \hi\ column density (\rhi). Our main results are summarized as follows: 

\begin{enumerate}
    \item CNNs, trained by realistic synthetic observations, are a promising tool for diagnosing ISM properties from spectral line observations. 
    
    \item By validating the trained CNN on a sample of observed $21\rm\,cm$ emission/absorption spectral line pairs \citep{heiles2003, murray2018b}, we demonstrate that the CNN predictions are equivalently, if not more, accurate than previous methods for quantifying \hi\ properties from $21\rm\,cm$ emission alone.
    
    \item At high Galactic latitudes, the CNM is ubiquitous but not the dominant \hi\ phase (\fcnm$<0.2$), in agreement with $21\rm\,cm$ absorption line studies. 
    
    \item The overall correction for optically-thick \hi\ to naive column density estimates is small ($\sim 5\%$ for $|b|>30^{\circ}$) but along individual lines of sight to CNM-rich structures it can be significant ($>20\%$). 
    
    \item Our results are fully consistent with the picture that filamentary \hi\ velocity structures aligned with the local magnetic field are caused by real density structures dominated by the CNM (Figure~\ref{f:fir_ratio}). 
    
    \item We find that \hi\ optical depth significantly affects the empirical correlation between \hi\ column density and dust reddening used to generate large-area maps of \ebv\ (Figure~\ref{f:ebv_ratio}), even in the diffuse, high-latitude regime studied here. Future efforts to improve \ebv\ maps in aid of cosmological surveys should leverage this result to increase fidelity. 
\end{enumerate}


\acknowledgements{
We thank the referee for their thoughtful report which has improved this work. 
We also thank John F. Wu for providing insightful comments on this manuscript.
C.E.M. is supported by an NSF Astronomy and Astrophysics Postdoctoral Fellowship under award AST-1801471.
The work of C.-G.K. was partly supported by a grant from the Simons Foundation (CCA 528307, E.C.O.) and NASA ATP grant NNX17AG26G. This work took part under the program Milky-Way-Gaia of the PSI2 project funded by the IDEX Paris-Saclay, ANR-11-IDEX-0003-02.

We acknowledge the Deep Skies Lab as a community of multi-domain experts and collaborators who have facilitated an environment of open discussion, idea-generation, and collaboration. This community was important for the development of this project. This research has made use of NASA's Astrophysics Data System. 

This work makes use of data from the Karl G. Jansky Very Large Array, operated by the National Radio Astronomy Observatory (NRAO). NRAO is a facility of the NSF operated under cooperative agreement by Associated Universities, Inc. The Arecibo Observatory is operated by SRI International under a cooperative agreement with the National Science Foundation (AST-1100968), and in alliance with Ana G. M\'{e}ndez-Universidad Metropolitana, and the Universities Space Research Association.}

\software{Astropy \citep{astropy2013}, NumPy \citep{vanderwalt2011}, matplotlib \citep{hunter2007}, glue \citep{beaumont2015}, Keras \citep{chollet2015}, Tensorflow \citep{tensorflow2015}, scipy \citep{scipy2020}, GaussPy \citep{lindner2015}, GaussPy+ \citep{riener2019}, pymc3 \citep{pymc3}}

\bibliography{ms}

\appendix

\section{Uncertainty Maps}
\label{ap:uncert_maps}

In this Appendix we present the all-Arecibo sky (at $|b|>30^{\circ}$) maps of the uncertainty in \fcnm\ and \rhi\ generated by the CNN ($\sigma_{f_{\rm CNM, CNN}}$ and $\sigma_{\mathcal{R}_{\rm HI,CNN}}$; Section~\ref{sec:uncertainty}). These uncertainties represent the contribution from the trained parameters (e.g., the weights and biases connecting the network layers). As a result, they do not incorporate the uncertainties in the network architecture or training dataset, and are therefore lower limits to the true uncertainty. We emphasize that the true uncertainties are likely dominated by the fact that the training dataset is not perfectly representative of real observations (as discussed in Section~\ref{sec:results}). We note that more sophisticated tools \citep[e.g., concrete dropout;][]{gal2017} have been shown to give better-calibrated uncertainty estimates, and will be explored in future work. 

\begin{figure*}
\begin{center}
\includegraphics[width=\textwidth]{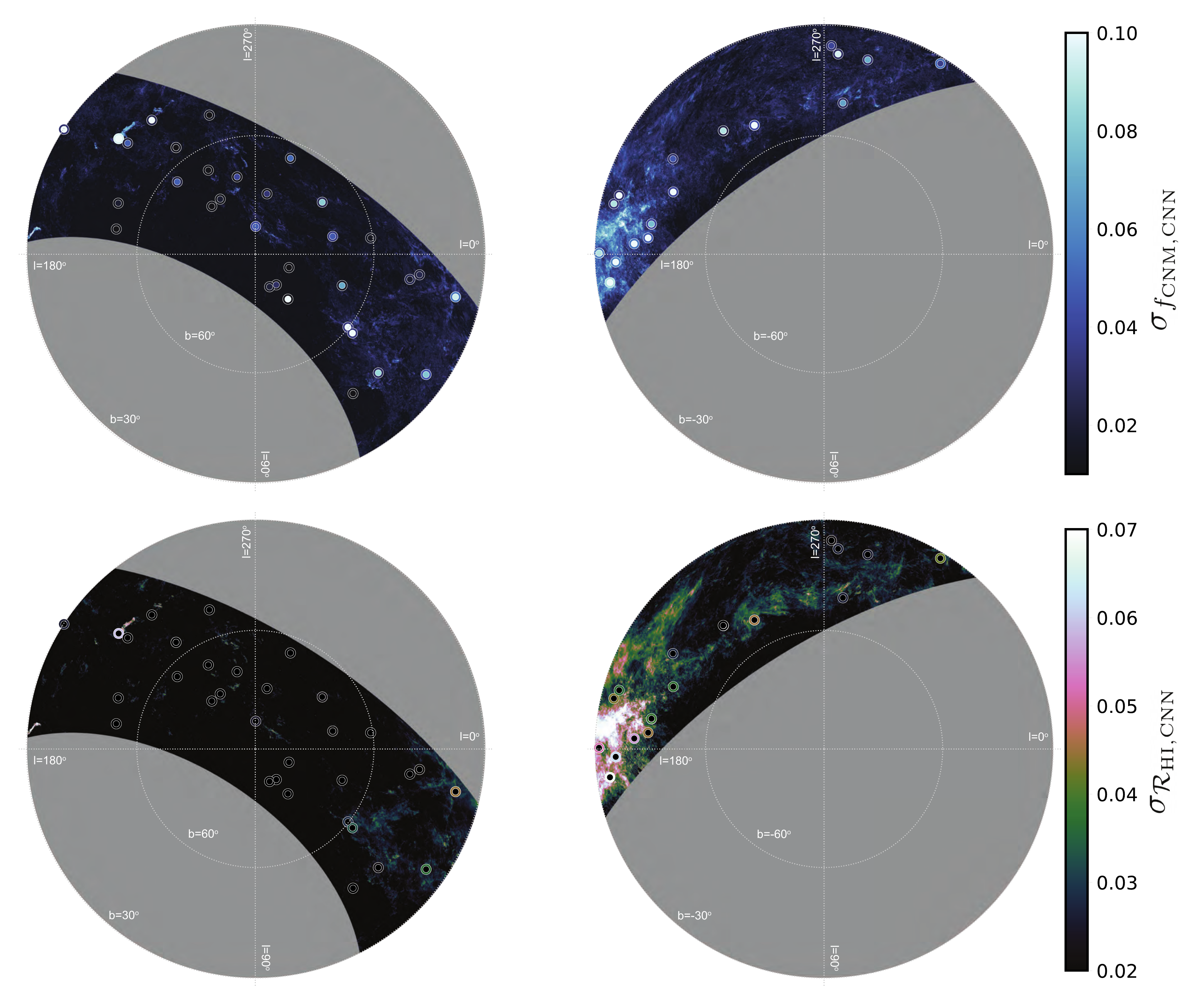}
\caption{Maps of the uncertainty in \fcnm\ (top) and \rhi\ (bottom) due to the trained CNN parameters, computed as the standard deviation following 25 trials. The observed LOS from the validation set are included as circles which are colored by the observed uncertainties (inside circles) and the CNN uncertainties (outside circles).}
\label{f:cnn_uncertainties}
\end{center}
\end{figure*}

\begin{figure*}
\begin{center}
\includegraphics[width=\textwidth]{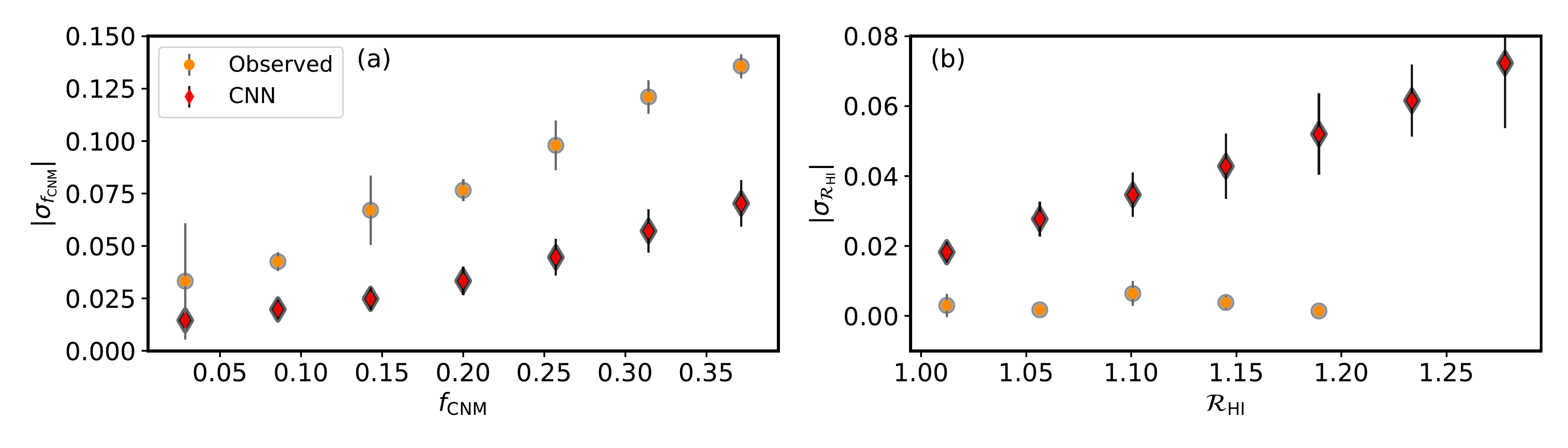}
\caption{Comparison of the mean empirical uncertainty ($|\sigma|$) in bins of \fcnm\ (a) and \rhi\ (b) for the CNN model of GALFA-\hi\ (red diamonds) and the \nabs\ absorption LOS of the validation set (orange circles). The error bars represent the standard deviations within each bin.}
\label{f:uncertainty_comparison}
\end{center}
\end{figure*}

To compare the uncertainties of the CNN predictions and the observed constraints, in Figure~\ref{f:uncertainty_comparison} we plot the mean uncertainties in bins of  \fcnm\ (a) and \rhi\ (b). Given the systematic uncertainty in Equation~\ref{e:fcnm}, including setting the values of $T_c$ and $T_{s,w}$, as well as the increased uncertainty in the observed sample at low-\fcnm\ (Figure~\ref{f:uncertainty_comparison}), our estimates for \fcnm$<0.1$ are typically uncertain. Furthermore, as discussed above, the uncertainties on the CNN predictions in this range represent lower limits. We observe that with increasing \fcnm\ and \rhi, the complexity of the spectral line structure increases, resulting in larger uncertainties in both the observed and CNN estimates. We are therefore most confident in our predictions for \fcnm\ in a the intermediate range of $0.1\lesssim$\fcnm$\lesssim0.2$ in this work. 

\begin{figure*}
\begin{center}
\includegraphics[width=0.98\textwidth]{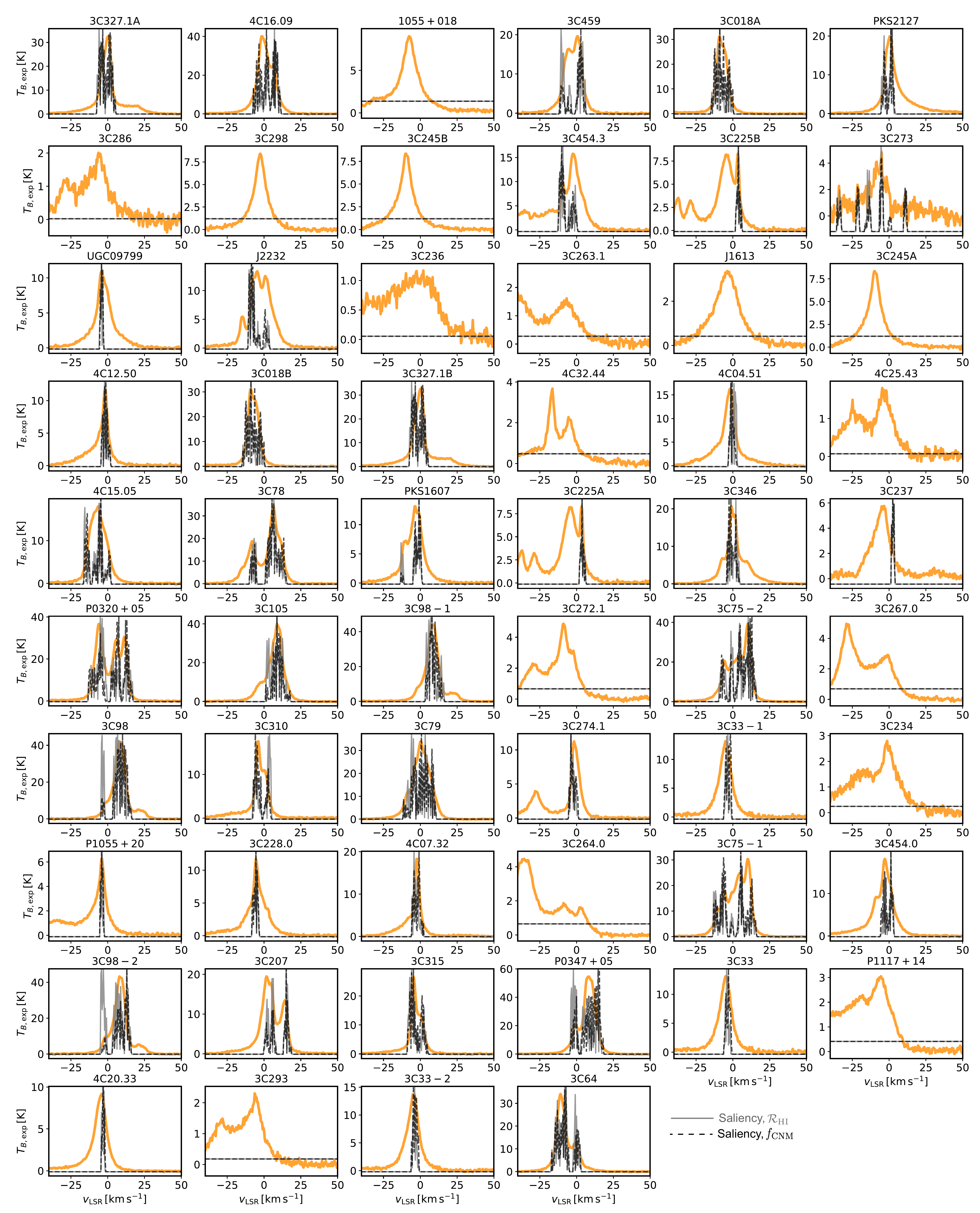}
\caption{Illustration of the salient components of the observed validation spectra (orange) which drive the predictions of \fcnm\ and \rhi. We compute the ``saliency" (see Section~\ref{sec:saliency}) of the final fully-connected output layer for \fcnm\ (grey) and \rhi\ (black dashed). We find consistent results between the observed and synthetic samples. The saliency spectra are arbitrarily scaled. }
\label{f:sal_maps_all}
\end{center}
\end{figure*}

\section{Saliency Spectra}
\label{ap:sal}

In this section, we display saliency spectra corresponding to the final fully-connected layer of the CNN (Figure~\ref{f:architecture}) for the \nabs\ LOS. In Figure~\ref{f:sal_maps_all}, we include the observed \tbexp\ spectra and the saliency spectra for \fcnm\ and \rhi\ (see Section~\ref{sec:saliency} for more details).

\section{Gaussian Decomposition with GaussPy+}
\label{ap:gp}

As a point of comparison for the CNN results, we decompose the GALFA-\hi\ $21\rm\,cm$ observations into Gaussian functions. By nature, any Gaussian decomposition will not present a unique solution. So, to eliminate the need for subjective input for the decomposition, we use the Autonomous Gaussian Decomposition algorithm \citep[AGD;][]{lindner2015}. AGD determines the number of Gaussian components and their properties for autonomously and simultaneously by analyzing the numerical derivatives of spectral lines, which requires regularization in the presence of spectral noise. The regularization parameters are determined via supervised machine learning on a training set with known decomposition parameters. The open-development package GaussPy\footnote{http://github.com/gausspy/gausspy} is the Python implementation of the AGD algorithm. GaussPy has been used to decompose $21\rm\,cm$ emission and absorption lines probing the local ISM \citep{murray2017,murray2018b,denes2018}.

To perform the decomposition of the GALFA-\hi\ data cubes, we use GaussPy+\footnote{http://github.com/mriener/gausspyplus} \citep{riener2019}, an open-source Python package which builds on GaussPy. GaussPy+ streamlines the implementation of AGD and also introduces new parameters for decomposing data cubes by imposing continuity between neighboring pixel solutions, and re-fitting in complex or un-physical regimes \citep[e.g., negative components, highly-blended features; see][]{riener2019}. For our application here, we use GaussPy+ for its logistical upgrades to GaussPy and do not implement the additional spatial re-fitting steps. We plan to conduct a detailed study of the effects of spatial re-fitting on the accuracy of \hi\ spectral cube decomposition in future work.

To conduct a comparison with the CNN results, rather than decompose the full GALFA-\hi\ sky, we extract $9\times9$ pixel sub-cubes centered on each of the \nabs\ absorption targets (e.g., position-position-velocity datasets). We also extract a larger sub-cube surrounding the Local Leo Cold Cloud \citep{verschuur1969,peek2011b} as a test case (Figure~\ref{f:llcc}). We prepare the data for GaussPy+ by extracting spectra from all sub-cubes and estimating their uncertainties as a function of velocity by computing the RMS noise in offline channels via helper functions implemented in GaussPy+. Next, we determine the appropriate regularization parameters for conducting ``two-phase" decomposition, which is designed to capture spectral features from both the CNM and WNM \citep[i.e., narrow and broad;][]{lindner2015}. Rather than re-train AGD with GaussPy+, we use the two-phase trained regularization parameters adopted by \citet{murray2017} in their study of the 21-SPONGE and KOK14 spectral libraries: $\alpha_1=1.17$ and $\alpha_2=3.75$. These values are suitable to apply to the GALFA-\hi\ sub-cubes, as they were trained using KOK14 data for the purpose of analyzing \tbexp\ from GALFA-\hi\ as part of the 21-SPONGE survey. With these regularization parameters in hand, we decompose all spectra from the selected sub-cubes using GaussPy+. As noted above, we halt the decomposition after its first iteration (i.e., after all spectra are decomposed independently). In future work, we will analyze the effect of subsequent iterations of spatial re-fitting \citep[e.g.,][]{riener2019, riener2020} on derived \hi\ properties.

\begin{figure*}
\begin{center}
\includegraphics[width=0.98\textwidth]{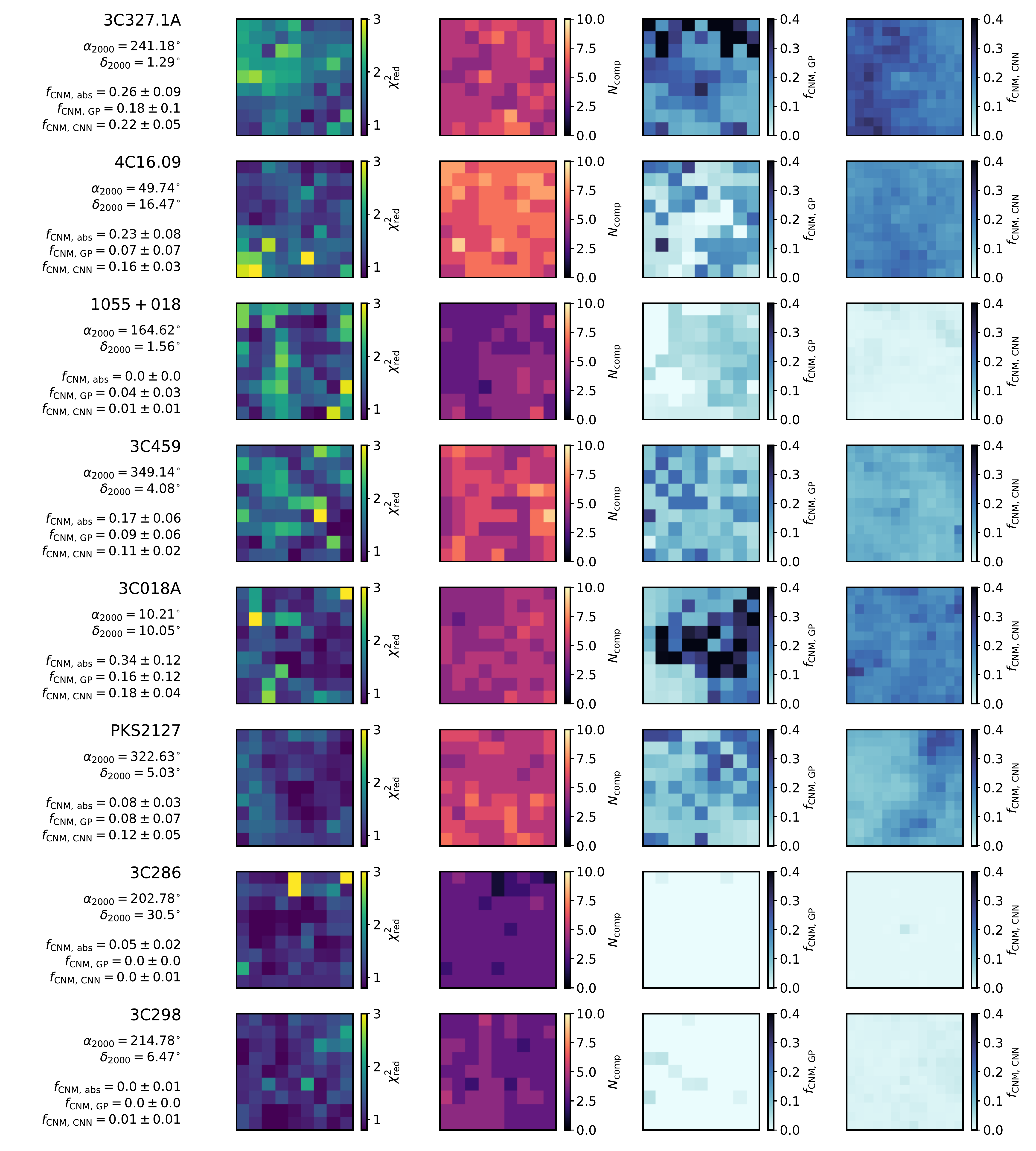}
\caption{Illustration of the GaussPy+ decomposition of $9\times9$ pixel GALFA-\hi\ sub-cubes ($9^{\prime}\times9^{\prime}$ centered on each of the \nabs\ absorption targets in the validation sample. For each target, we list the source name, coordinates ($\alpha_{2000}$, $\delta_{2000}$) and three estimates of \fcnm\ for this region of sky, including directly from $21\rm\,cm$ optical depth measurements (\fcnm$_{,\,\rm abs}$), from the CNN (\fcnm$_{,\,\rm CNN}$) and from GaussPy+ (\fcnm$_{,\,\rm GP}$). From left to right, we include maps of $\chi^2_{\rm red}$ and the number of components from the GaussPy+ decomposition ($N_{\rm comp}$), and maps of \fcnm$_{,\,\rm GP}$ and \fcnm$_{,\,\rm CNN}$. (maps for all sources will be available in the online journal).}
\label{f:gp_maps}
\end{center}
\end{figure*}

By inspection of the decomposition results, we observe that GaussPy+ performs well. In Figure~\ref{f:gp_maps} we display maps of the reduced chi-squared ($\chi^2_{\rm red}$) for the GaussPy+ fit to each pixel to illustrate its performance. 

To estimate \fcnm\ from the Gaussian decomposition results, for each region we extract all components with CNM-like temperatures. We define these components as having widths (measured by the full width at half maximum) $\rm FWHM<7\rm\,km\,s^{-1}$ (equivalent to line width $\sigma<3\rm\,km\,s^{-1}$), which corresponds to maximum kinetic temperatures (i.e., including thermal and non-thermal broadening) of $T_{k,\rm max}=21.866\times {\rm FWHM}^2 \lesssim 1000\rm\,K$). We selected this CNM cutoff line width to be generally consistent with previous work \citep[e.g.,][]{takakubo1967, kalberla2018, marchal2019}.

For the \nabs\ sub-cubes surrounding the absorption targets, we compute the expected \fcnm\ values as the median of the 72 values in a $9\times9$ pixel grid surrounding each source, excluding the central $3\times3$ pixels which are typically contaminated by absorption structure (e.g., consistent with the computation of \tbexp). In Figure~\ref{f:gp_maps} we compare the \fcnm\ maps from GaussPy+ with the corresponding \fcnm\ maps from the CNN. For the LLCC sub-cube, we present the full \fcnm\ map from GaussPy+ in Figure~\ref{f:llcc}(c). We note that the CNN \fcnm\ maps are remarkably smooth, and appear even smoother than the GaussPy+ maps, despite the absence of spatial smoothing information in both methods. The effect of including smoothing parameters in new applications of GaussPy+ to \hi\ data will be explored in future work. We also note that this smoothness may contribute to reducing the empirical uncertainties in the CNN-derived parameters.

\figsetstart
\figsetnum{17}
\figsettitle{Comparison with Gaussian Decomposition}

\figsetgrpstart
\figsetgrpnum{17.1}
\figsetgrptitle{GaussPy+/CNN comparison}
\figsetplot{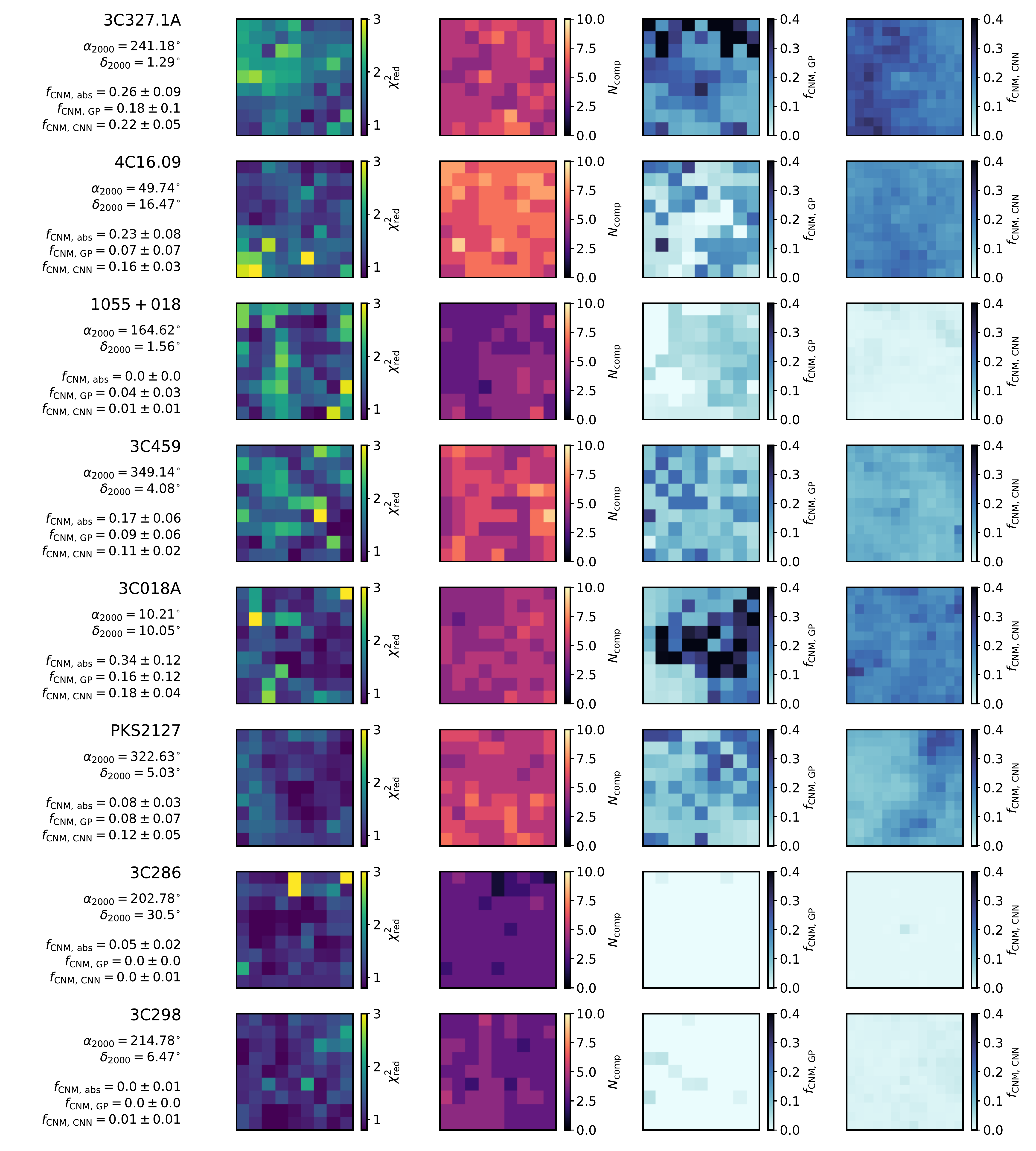}
\figsetgrpnote{Illustration of the GaussPy+ decomposition of $9\times9$ pixel GALFA-\hi\ sub-cubes ($9^{\prime}\times9^{\prime}$ centered on each of the \nabs\ absorption targets in the validation sample. For each target, we list the source name, coordinates ($\alpha_{2000}$, $\delta_{2000}$) and three estimates of \fcnm\ for this region of sky, including directly from $21\rm\,cm$ optical depth measurements (\fcnm$_{,\,\rm abs}$), from the CNN (\fcnm$_{,\,\rm CNN}$) and from GaussPy+ (\fcnm$_{,\,\rm GP}$). From left to right, we include maps of $\chi^2_{\rm red}$ and the number of components from the GaussPy+ decomposition ($N_{\rm comp}$), and maps of \fcnm$_{,\,\rm GP}$ and \fcnm$_{,\,\rm CNN}$.}
\figsetgrpend

\figsetgrpstart
\figsetgrpnum{17.2}
\figsetgrptitle{GaussPy+/CNN comparison (contd)}
\figsetplot{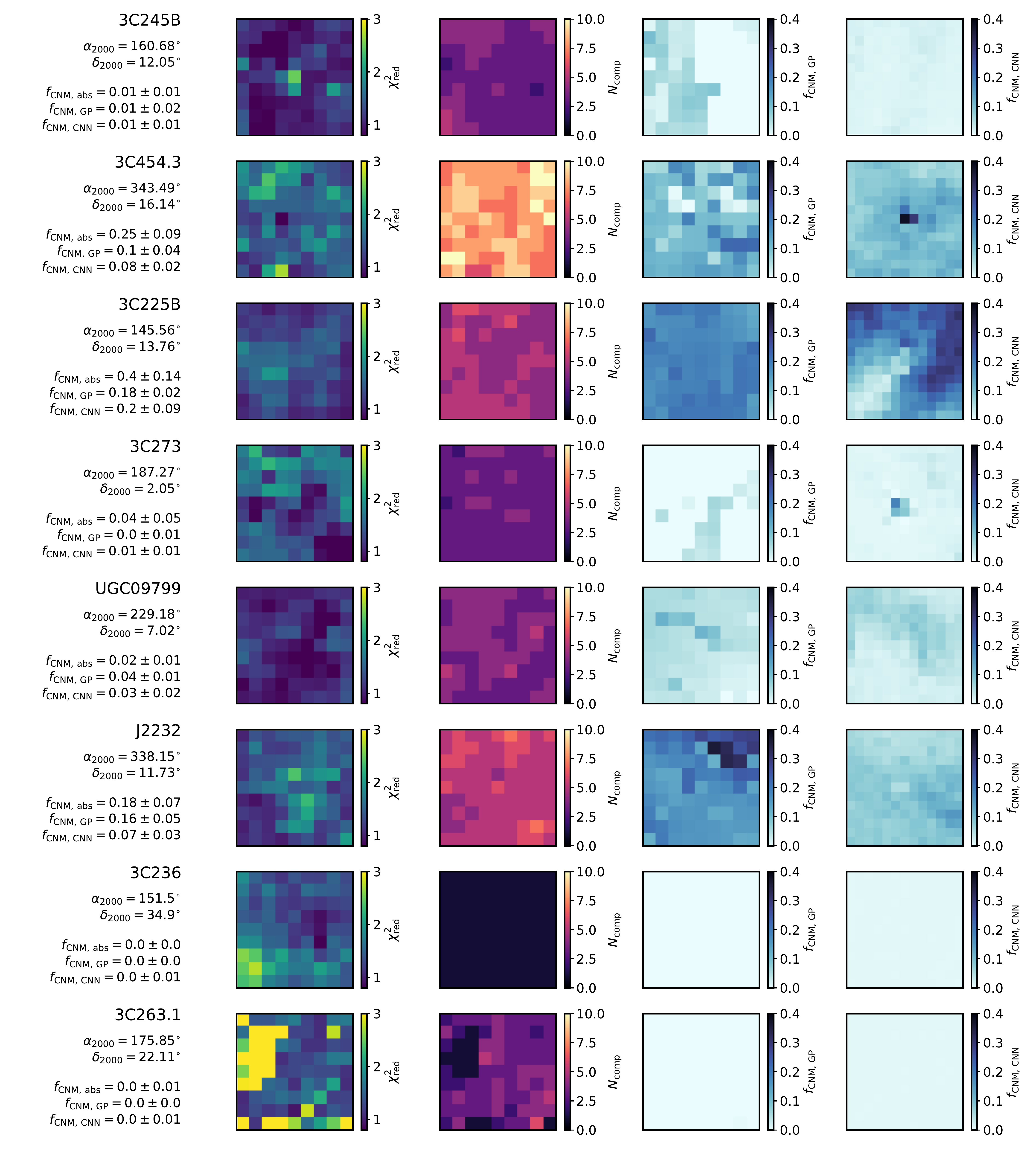}
\figsetgrpnote{Illustration of the GaussPy+ decomposition of $9\times9$ pixel GALFA-\hi\ sub-cubes ($9^{\prime}\times9^{\prime}$ centered on each of the \nabs\ absorption targets in the validation sample. For each target, we list the source name, coordinates ($\alpha_{2000}$, $\delta_{2000}$) and three estimates of \fcnm\ for this region of sky, including directly from $21\rm\,cm$ optical depth measurements (\fcnm$_{,\,\rm abs}$), from the CNN (\fcnm$_{,\,\rm CNN}$) and from GaussPy+ (\fcnm$_{,\,\rm GP}$). From left to right, we include maps of $\chi^2_{\rm red}$ and the number of components from the GaussPy+ decomposition ($N_{\rm comp}$), and maps of \fcnm$_{,\,\rm GP}$ and \fcnm$_{,\,\rm CNN}$.}
\figsetgrpend

\figsetgrpstart
\figsetgrpnum{17.3}
\figsetgrptitle{GaussPy+/CNN comparison (contd)}
\figsetplot{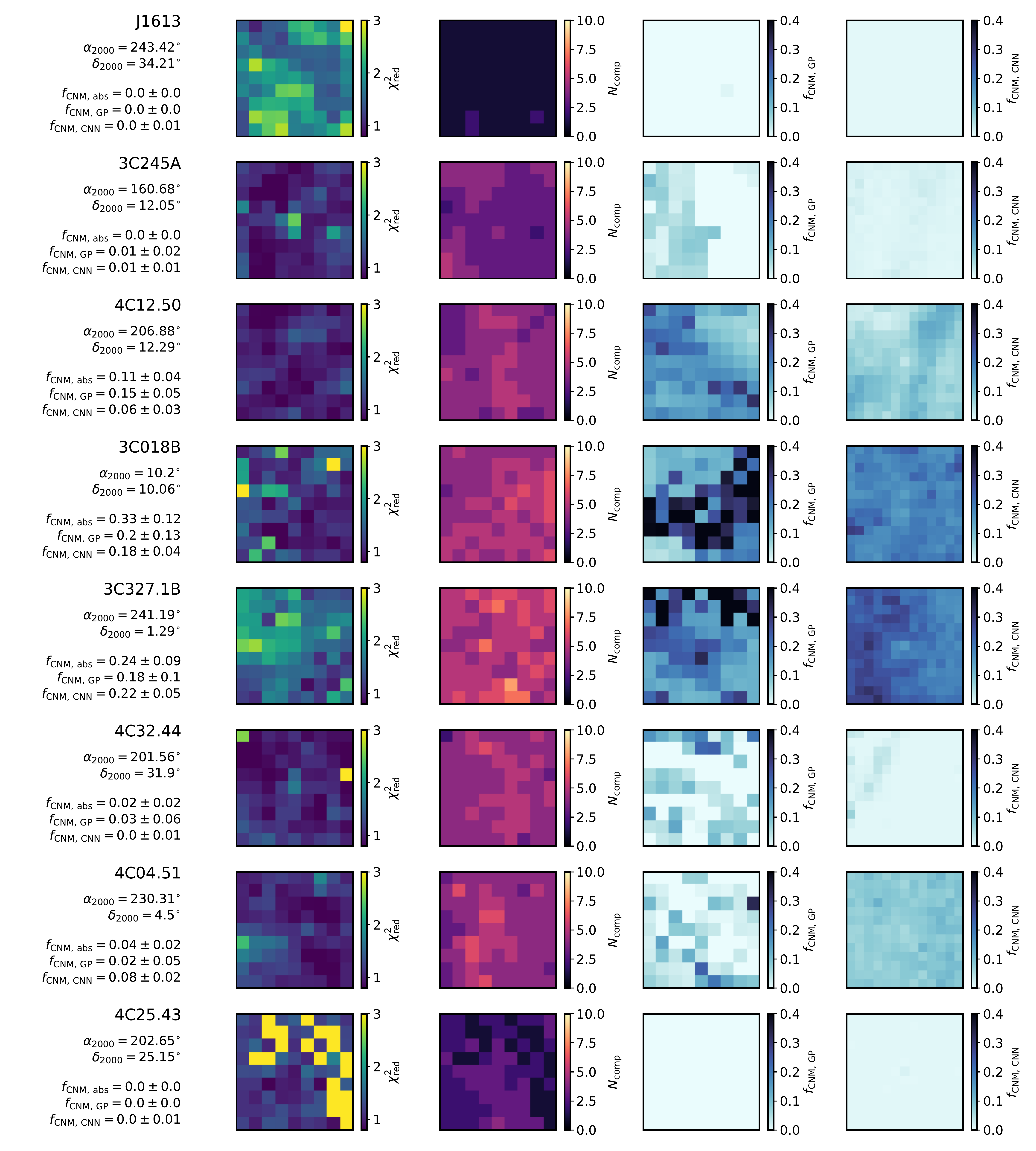}
\figsetgrpnote{Illustration of the GaussPy+ decomposition of $9\times9$ pixel GALFA-\hi\ sub-cubes ($9^{\prime}\times9^{\prime}$ centered on each of the \nabs\ absorption targets in the validation sample. For each target, we list the source name, coordinates ($\alpha_{2000}$, $\delta_{2000}$) and three estimates of \fcnm\ for this region of sky, including directly from $21\rm\,cm$ optical depth measurements (\fcnm$_{,\,\rm abs}$), from the CNN (\fcnm$_{,\,\rm CNN}$) and from GaussPy+ (\fcnm$_{,\,\rm GP}$). From left to right, we include maps of $\chi^2_{\rm red}$ and the number of components from the GaussPy+ decomposition ($N_{\rm comp}$), and maps of \fcnm$_{,\,\rm GP}$ and \fcnm$_{,\,\rm CNN}$.}
\figsetgrpend

\figsetgrpstart
\figsetgrpnum{17.4}
\figsetgrptitle{GaussPy+/CNN comparison (contd)}
\figsetplot{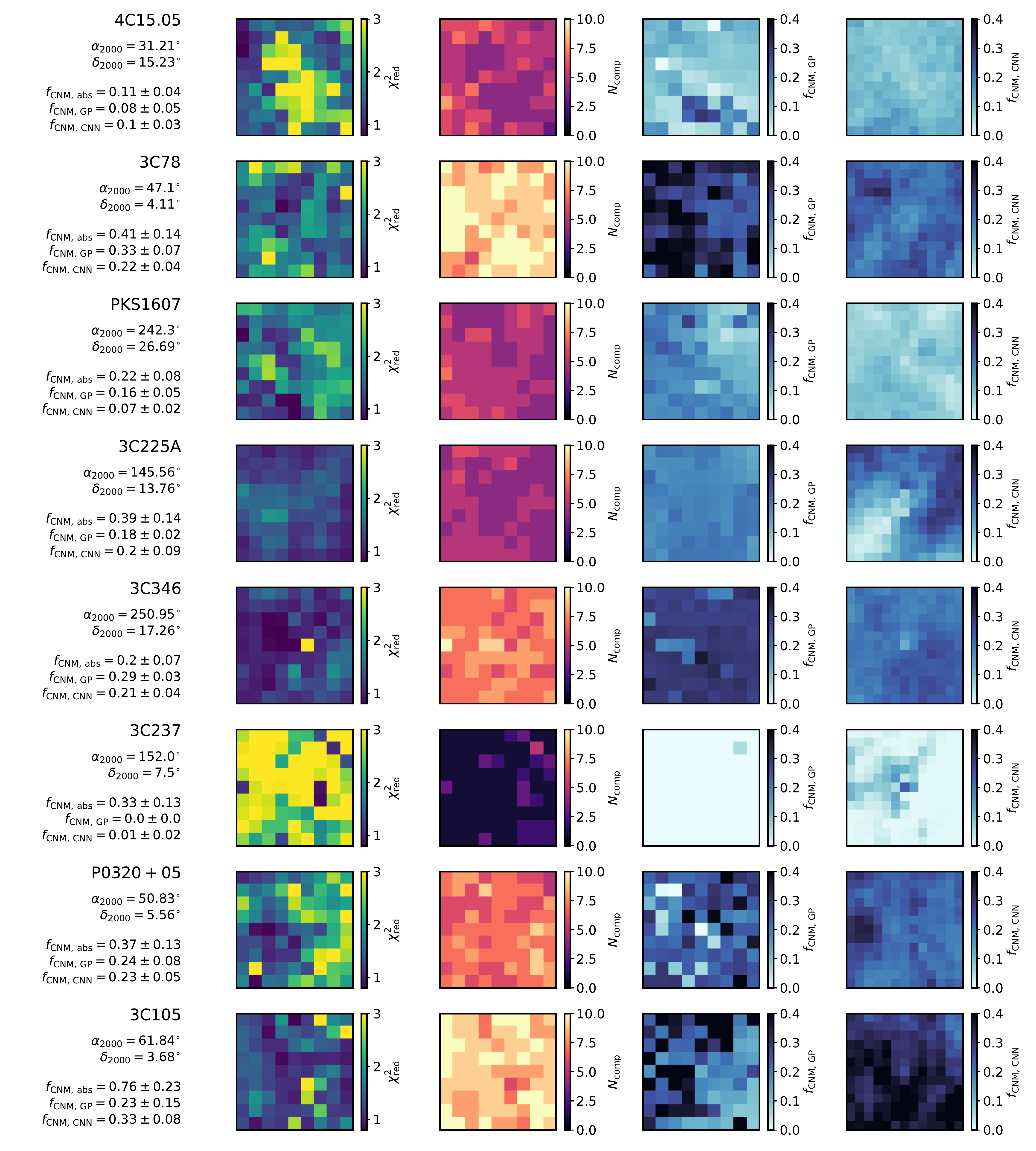}
\figsetgrpnote{Illustration of the GaussPy+ decomposition of $9\times9$ pixel GALFA-\hi\ sub-cubes ($9^{\prime}\times9^{\prime}$ centered on each of the \nabs\ absorption targets in the validation sample. For each target, we list the source name, coordinates ($\alpha_{2000}$, $\delta_{2000}$) and three estimates of \fcnm\ for this region of sky, including directly from $21\rm\,cm$ optical depth measurements (\fcnm$_{,\,\rm abs}$), from the CNN (\fcnm$_{,\,\rm CNN}$) and from GaussPy+ (\fcnm$_{,\,\rm GP}$). From left to right, we include maps of $\chi^2_{\rm red}$ and the number of components from the GaussPy+ decomposition ($N_{\rm comp}$), and maps of \fcnm$_{,\,\rm GP}$ and \fcnm$_{,\,\rm CNN}$.}
\figsetgrpend

\figsetgrpstart
\figsetgrpnum{17.5}
\figsetgrptitle{GaussPy+/CNN comparison (contd)}
\figsetplot{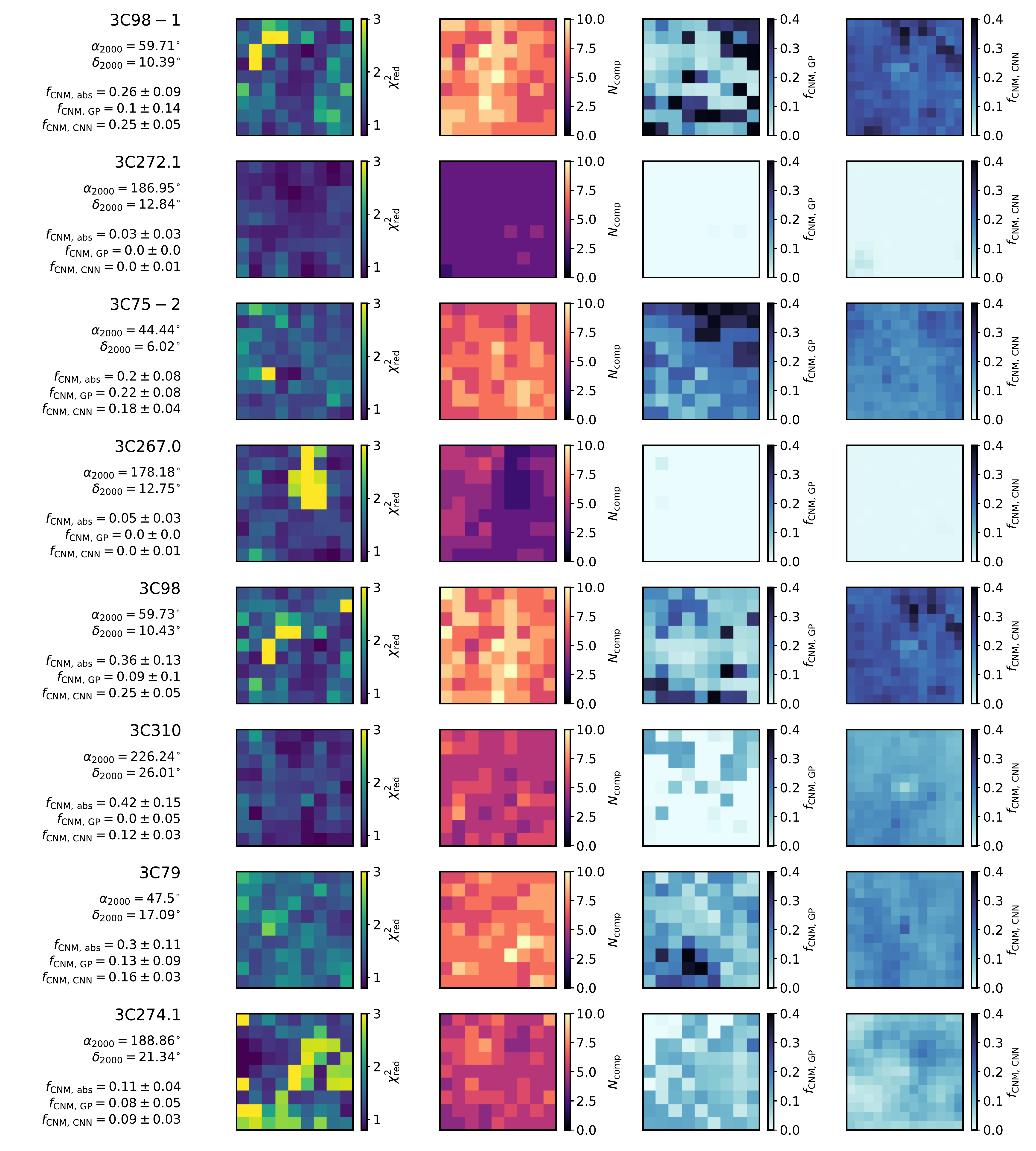}
\figsetgrpnote{Illustration of the GaussPy+ decomposition of $9\times9$ pixel GALFA-\hi\ sub-cubes ($9^{\prime}\times9^{\prime}$ centered on each of the \nabs\ absorption targets in the validation sample. For each target, we list the source name, coordinates ($\alpha_{2000}$, $\delta_{2000}$) and three estimates of \fcnm\ for this region of sky, including directly from $21\rm\,cm$ optical depth measurements (\fcnm$_{,\,\rm abs}$), from the CNN (\fcnm$_{,\,\rm CNN}$) and from GaussPy+ (\fcnm$_{,\,\rm GP}$). From left to right, we include maps of $\chi^2_{\rm red}$ and the number of components from the GaussPy+ decomposition ($N_{\rm comp}$), and maps of \fcnm$_{,\,\rm GP}$ and \fcnm$_{,\,\rm CNN}$.}
\figsetgrpend

\figsetgrpstart
\figsetgrpnum{17.6}
\figsetgrptitle{GaussPy+/CNN comparison (contd)}
\figsetplot{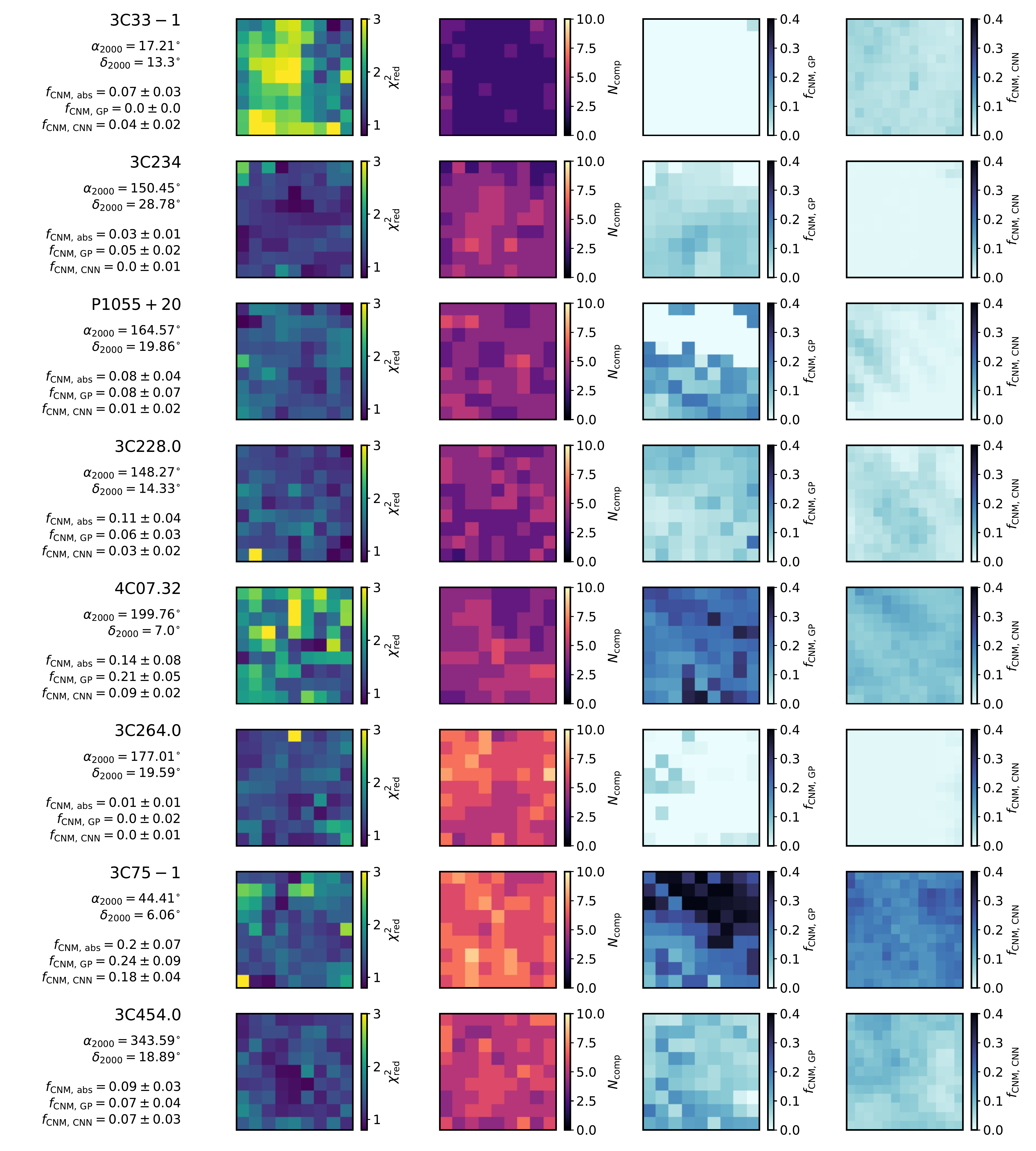}
\figsetgrpnote{Illustration of the GaussPy+ decomposition of $9\times9$ pixel GALFA-\hi\ sub-cubes ($9^{\prime}\times9^{\prime}$ centered on each of the \nabs\ absorption targets in the validation sample. For each target, we list the source name, coordinates ($\alpha_{2000}$, $\delta_{2000}$) and three estimates of \fcnm\ for this region of sky, including directly from $21\rm\,cm$ optical depth measurements (\fcnm$_{,\,\rm abs}$), from the CNN (\fcnm$_{,\,\rm CNN}$) and from GaussPy+ (\fcnm$_{,\,\rm GP}$). From left to right, we include maps of $\chi^2_{\rm red}$ and the number of components from the GaussPy+ decomposition ($N_{\rm comp}$), and maps of \fcnm$_{,\,\rm GP}$ and \fcnm$_{,\,\rm CNN}$.}
\figsetgrpend

\figsetgrpstart
\figsetgrpnum{17.7}
\figsetgrptitle{GaussPy+/CNN comparison (contd)}
\figsetplot{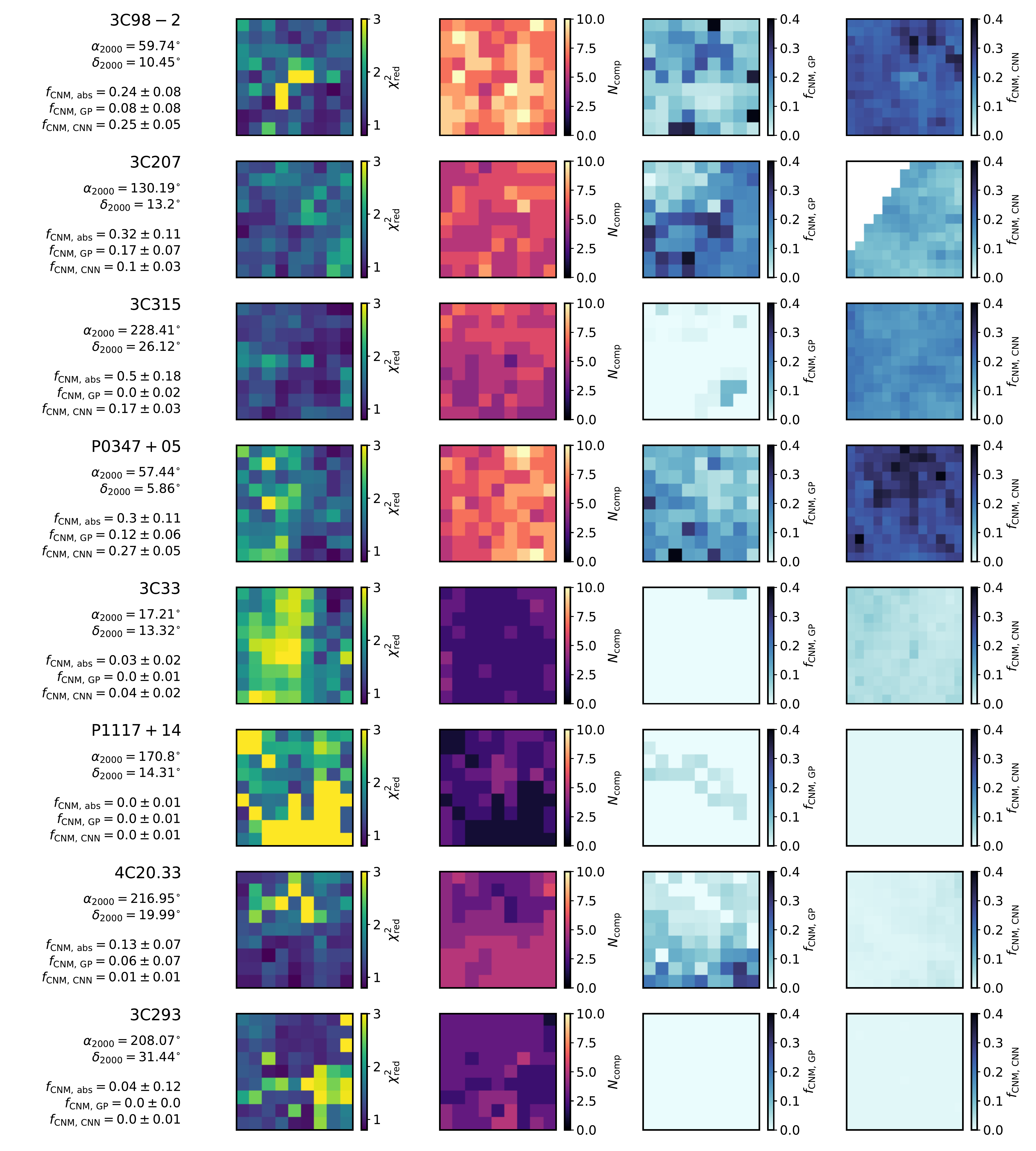}
\figsetgrpnote{Illustration of the GaussPy+ decomposition of $9\times9$ pixel GALFA-\hi\ sub-cubes ($9^{\prime}\times9^{\prime}$ centered on each of the \nabs\ absorption targets in the validation sample. For each target, we list the source name, coordinates ($\alpha_{2000}$, $\delta_{2000}$) and three estimates of \fcnm\ for this region of sky, including directly from $21\rm\,cm$ optical depth measurements (\fcnm$_{,\,\rm abs}$), from the CNN (\fcnm$_{,\,\rm CNN}$) and from GaussPy+ (\fcnm$_{,\,\rm GP}$). From left to right, we include maps of $\chi^2_{\rm red}$ and the number of components from the GaussPy+ decomposition ($N_{\rm comp}$), and maps of \fcnm$_{,\,\rm GP}$ and \fcnm$_{,\,\rm CNN}$.}
\figsetgrpend

\figsetgrpstart
\figsetgrpnum{17.8}
\figsetgrptitle{GaussPy+/CNN comparison (contd)}
\figsetplot{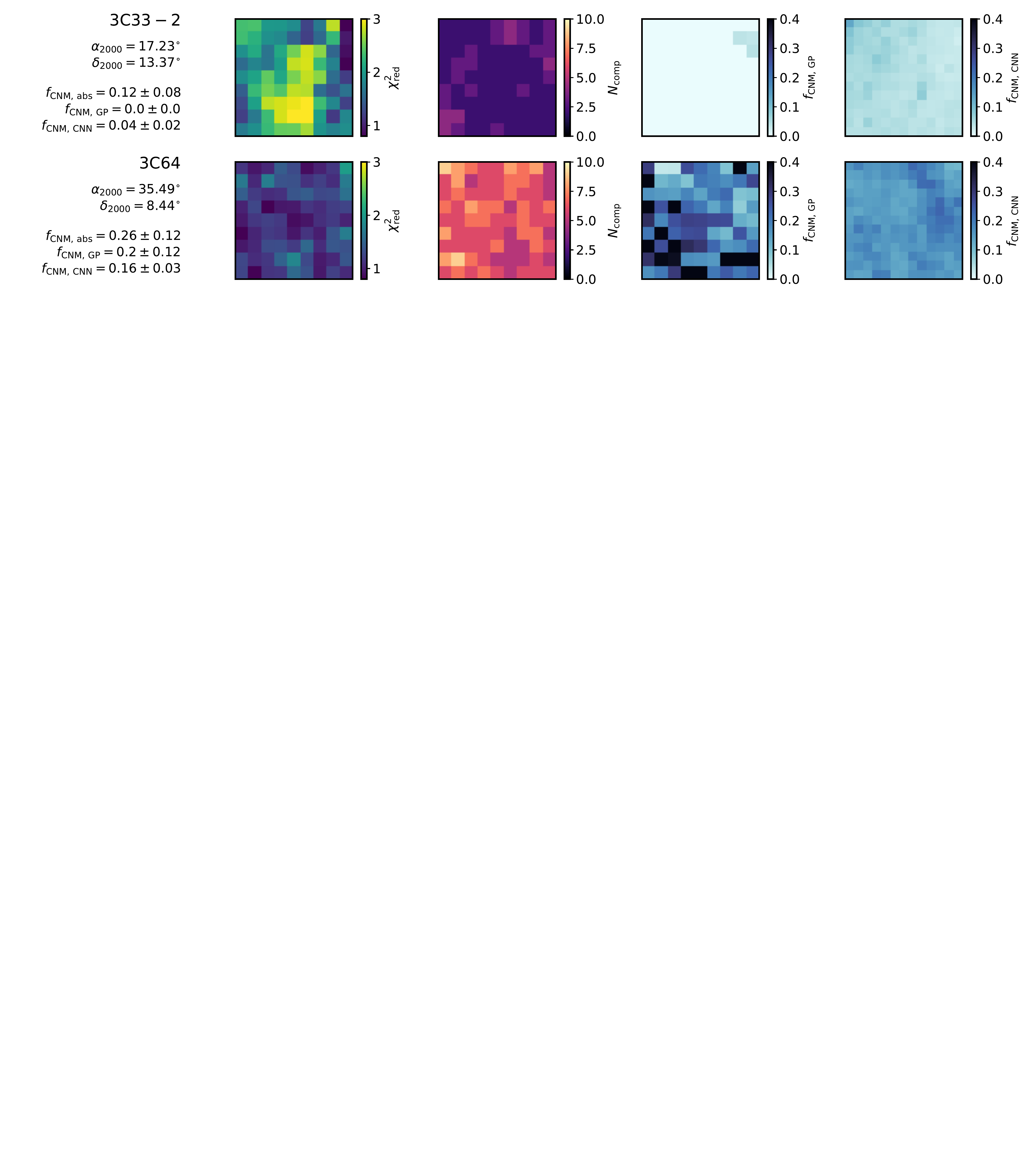}
\figsetgrpnote{Illustration of the GaussPy+ decomposition of $9\times9$ pixel GALFA-\hi\ sub-cubes ($9^{\prime}\times9^{\prime}$ centered on each of the \nabs\ absorption targets in the validation sample. For each target, we list the source name, coordinates ($\alpha_{2000}$, $\delta_{2000}$) and three estimates of \fcnm\ for this region of sky, including directly from $21\rm\,cm$ optical depth measurements (\fcnm$_{,\,\rm abs}$), from the CNN (\fcnm$_{,\,\rm CNN}$) and from GaussPy+ (\fcnm$_{,\,\rm GP}$). From left to right, we include maps of $\chi^2_{\rm red}$ and the number of components from the GaussPy+ decomposition ($N_{\rm comp}$), and maps of \fcnm$_{,\,\rm GP}$ and \fcnm$_{,\,\rm CNN}$.}
\figsetgrpend

\figsetend

\begin{figure*}
\begin{center}
\includegraphics[width=\textwidth]{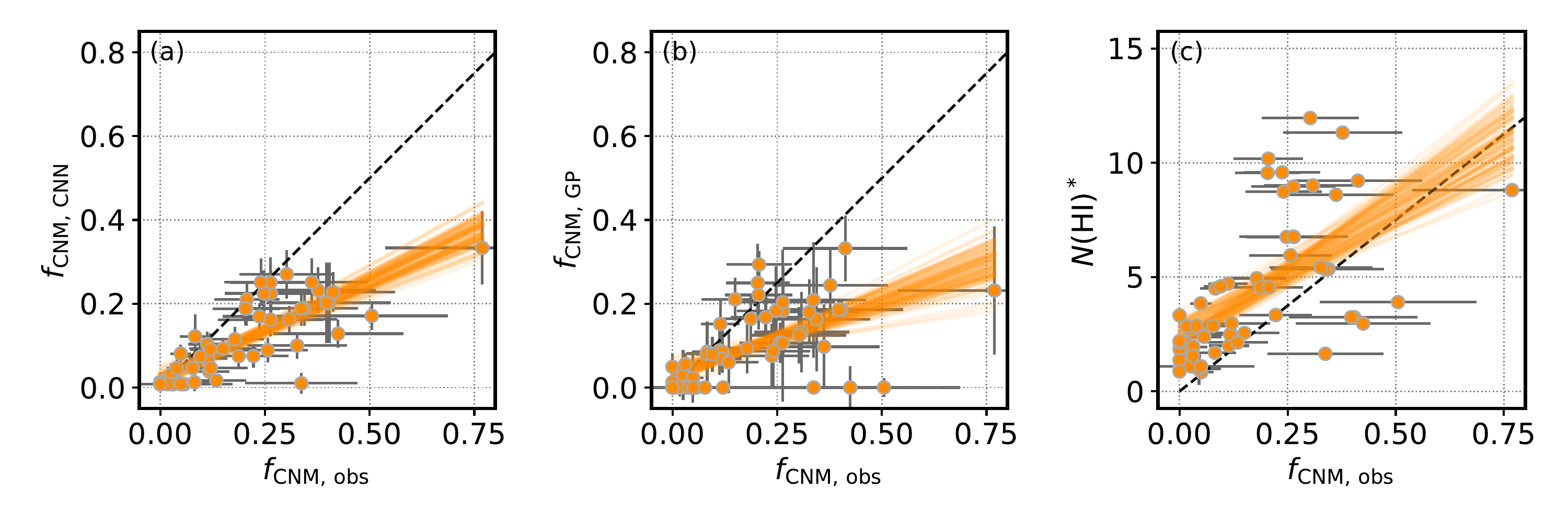}
\caption{Comparison between \fcnm\ constraints from $21\rm\,cm$ absorption observations and \fcnm\ inferred by the CNN (a) GaussPy+ (b) and \nhithin\ (c). For each tracer, we fit a Bayesian linear model and include fits constructed from 100 draws of the posterior distributions for slope and intercept. We find that the CNN performs better at reproducing \fcnm\ traced by $21\rm\,cm$ absorption observations than the Gaussian decomposition method and the \nhithin\ model. }
\label{f:gauss_compare}
\end{center}
\end{figure*}

In Figure~\ref{f:gauss_compare} we compare the ``ground truth" constraints from $21\rm\,cm$ absorption measurements with the results of the CNN (a), and Gaussian decomposition (b). By eye, we conclude that the CNN performs just as well, if not better, at estimating \fcnm\ as the GaussPy+ method. To quantify this, we construct a simple Bayesian linear regression model for each method. Using pymc3 \citep{pymc3}, we impose normal prior distributions for the slope and intercept of the linear model, and perform Markov Chain Monte Carlo (MCMC) with 2000 draws. From the posterior distributions for the slope and intercept of each fit, we draw 100 samples and include these fits in Figure~\ref{f:gauss_compare}. Although neither case is consistent with 1:1, based on the mean and standard deviation ($\sigma_{\rm slope}$) of the posterior distributions, the slope in the CNN case is positive with higher significance ($8.5\sigma_{\rm slope}$) than the decomposition case ($4.6\sigma_{\rm slope}$), indicating a better correlation. In Figure~\ref{f:gauss_compare} we also include the results of predicting \fcnm$_{,\rm obs}$ by simply integrating $T_B(v)$ (i.e., \nhithin), and find that a positive slope with less significance than the CNN case, but similar to the decomposition case ($4.5\sigma_{\rm slope}$). 


Finally, we conduct a simple test to see if the CNN predictions are based on relevant spectral information (as has been extensively tested for the Gaussian decomposition method in the literature). For each of the \nabs\ observed LOS, we compare \fcnm$_{,\rm obs}$ with the CNN prediction for an LOS with similar \nhithin\ (defined to be within $5\times10^{19}\rm\,cm^{-2}$).  Although there is still a positive linear correlation ($4.1\sigma_{\rm slope}$) it is significantly worse than the CNN predictions for the original sample (Figure~\ref{f:gauss_compare}a), indicating that although the integrated intensity of emission (\nhithin) contributes, more sophisticated features of the line profile are crucial for maximizing the accuracy of the CNN prediction. This result agrees with the saliency analysis (Figure~\ref{f:saliency} and~\ref{f:sal_maps_all}) which demonstrated, e.g., that the CNN is sensitive to narrow spectral features when inferring the presence of the CNM.

\section{Source Information}
\label{ap:sources}

In the following appendix we include information describing the \nabs\ LOS with observed $21\rm\,cm$ optical depth (\thi). Table~\ref{tab:sources} lists the coordinates, and the relevant extracted \hi\ properties.  

\begin{longtable*}{llcc|ll|ll|ll} 
\colhead{Name} & \colhead{Ref.} & \colhead{RA} & \colhead{Dec} & \colhead{$N({\rm HI})_{\rm thin}$} & \colhead{$N({\rm HI})_{\rm iso}$}  & \colhead{$f_{\rm CNM}$}  &  \colhead{$f_{\rm CNM,\,CNN}$}  & \colhead{$\mathcal{R}_{\rm HI}$}  & \colhead{$\mathcal{R}_{\rm HI, \,CNN}$}   \\
\colhead{}  & \colhead{}    & \colhead{(deg)} & \colhead{(deg)} & \colhead{($10^{20} \rm cm^{-2}$)}   & \colhead{($10^{20} \rm cm^{-2}$)}  &      \colhead{}         &     \colhead{}     &         \colhead{}               &         \colhead{}                            \\ 
\colhead{(1)} & \colhead{(2)} & \colhead{(3)} & \colhead{(4)} & \colhead{(5)} & \colhead{(6)} & \colhead{(7)} & \colhead{(8)} & \colhead{(9)} & \colhead{(10)}  \\ 
\hline 
3C018B  &  1  &  10.206  &  10.064  &  5.34$\, \pm \,  $0.15  &  5.94$\, \pm \,  $0.15  &  0.34$\, \pm \,  $0.13  &  0.19$\, \pm \,  $0.04  &  1.11$\, \pm \,  $0.01  &  1.11$\, \pm \,  $0.05  \\ 
3C018A  &  1  &  10.211  &  10.051  &  5.34$\, \pm \,  $0.15  &  5.95$\, \pm \,  $0.15  &  0.34$\, \pm \,  $0.13  &  0.19$\, \pm \,  $0.04  &  1.11$\, \pm \,  $0.01  &  1.11$\, \pm \,  $0.05  \\ 
3C33-1  &  2  &  17.211  &  13.308  &  2.87$\, \pm \,  $0.24  &  2.89$\, \pm \,  $0.24  &  0.08$\, \pm \,  $0.04  &  0.05$\, \pm \,  $0.02  &  1.01$\, \pm \,  $0.00  &  1.00$\, \pm \,  $0.02  \\ 
3C33  &  2  &  17.219  &  13.327  &  2.87$\, \pm \,  $0.25  &  2.89$\, \pm \,  $0.25  &  0.04$\, \pm \,  $0.02  &  0.05$\, \pm \,  $0.02  &  1.01$\, \pm \,  $0.00  &  1.00$\, \pm \,  $0.02  \\ 
3C33-2  &  2  &  17.233  &  13.372  &  2.97$\, \pm \,  $0.26  &  3.02$\, \pm \,  $0.26  &  0.12$\, \pm \,  $0.09  &  0.05$\, \pm \,  $0.02  &  1.02$\, \pm \,  $0.00  &  1.00$\, \pm \,  $0.02  \\ 
4C15.05  &  1  &  31.210  &  15.236  &  4.72$\, \pm \,  $0.16  &  4.79$\, \pm \,  $0.13  &  0.11$\, \pm \,  $0.05  &  0.10$\, \pm \,  $0.03  &  1.02$\, \pm \,  $0.01  &  1.03$\, \pm \,  $0.03  \\ 
3C64  &  2  &  35.495  &  8.449  &  6.77$\, \pm \,  $0.12  &  7.23$\, \pm \,  $0.12  &  0.26$\, \pm \,  $0.13  &  0.16$\, \pm \,  $0.04  &  1.07$\, \pm \,  $0.00  &  1.07$\, \pm \,  $0.04  \\ 
3C75-1  &  2  &  44.411  &  6.064  &  9.56$\, \pm \,  $0.10  &  9.90$\, \pm \,  $0.10  &  0.20$\, \pm \,  $0.08  &  0.19$\, \pm \,  $0.04  &  1.04$\, \pm \,  $0.00  &  1.08$\, \pm \,  $0.04  \\ 
3C75-2  &  2  &  44.448  &  6.021  &  10.18$\, \pm \,  $0.09  &  10.46$\, \pm \,  $0.09  &  0.21$\, \pm \,  $0.08  &  0.19$\, \pm \,  $0.04  &  1.03$\, \pm \,  $0.00  &  1.08$\, \pm \,  $0.04  \\ 
3C78  &  1  &  47.109  &  4.111  &  9.22$\, \pm \,  $0.12  &  10.56$\, \pm \,  $0.07  &  0.41$\, \pm \,  $0.15  &  0.23$\, \pm \,  $0.05  &  1.15$\, \pm \,  $0.01  &  1.11$\, \pm \,  $0.04  \\ 
3C79  &  2  &  47.500  &  17.099  &  9.01$\, \pm \,  $0.08  &  9.71$\, \pm \,  $0.08  &  0.31$\, \pm \,  $0.11  &  0.16$\, \pm \,  $0.03  &  1.08$\, \pm \,  $0.00  &  1.07$\, \pm \,  $0.04  \\ 
4C16.09  &  1  &  49.741  &  16.476  &  9.58$\, \pm \,  $0.18  &  10.49$\, \pm \,  $0.11  &  0.24$\, \pm \,  $0.09  &  0.17$\, \pm \,  $0.03  &  1.10$\, \pm \,  $0.01  &  1.10$\, \pm \,  $0.04  \\ 
P0320+05  &  2  &  50.834  &  5.570  &  11.32$\, \pm \,  $0.24  &  12.77$\, \pm \,  $0.24  &  0.38$\, \pm \,  $0.14  &  0.23$\, \pm \,  $0.06  &  1.13$\, \pm \,  $0.00  &  1.12$\, \pm \,  $0.06  \\ 
P0347+05  &  2  &  57.445  &  5.861  &  11.96$\, \pm \,  $0.08  &  14.40$\, \pm \,  $0.08  &  0.30$\, \pm \,  $0.11  &  0.27$\, \pm \,  $0.06  &  1.20$\, \pm \,  $0.00  &  1.17$\, \pm \,  $0.06  \\ 
3C98-1  &  2  &  59.714  &  10.398  &  8.96$\, \pm \,  $0.09  &  9.92$\, \pm \,  $0.09  &  0.26$\, \pm \,  $0.10  &  0.25$\, \pm \,  $0.06  &  1.11$\, \pm \,  $0.00  &  1.15$\, \pm \,  $0.05  \\ 
3C98  &  2  &  59.730  &  10.436  &  8.60$\, \pm \,  $0.10  &  9.81$\, \pm \,  $0.10  &  0.36$\, \pm \,  $0.13  &  0.25$\, \pm \,  $0.06  &  1.14$\, \pm \,  $0.00  &  1.16$\, \pm \,  $0.05  \\ 
3C98-2  &  2  &  59.745  &  10.458  &  8.73$\, \pm \,  $0.08  &  9.68$\, \pm \,  $0.08  &  0.24$\, \pm \,  $0.09  &  0.25$\, \pm \,  $0.06  &  1.11$\, \pm \,  $0.00  &  1.16$\, \pm \,  $0.05  \\ 
3C105  &  2  &  61.843  &  3.690  &  8.80$\, \pm \,  $0.08  &  13.72$\, \pm \,  $0.08  &  0.77$\, \pm \,  $0.23  &  0.33$\, \pm \,  $0.09  &  1.56$\, \pm \,  $0.00  &  1.20$\, \pm \,  $0.07  \\ 
3C207  &  2  &  130.199  &  13.207  &  5.42$\, \pm \,  $0.09  &  5.72$\, \pm \,  $0.09  &  0.33$\, \pm \,  $0.12  &  0.10$\, \pm \,  $0.03  &  1.06$\, \pm \,  $0.00  &  1.03$\, \pm \,  $0.03  \\ 
3C225A  &  1  &  145.564  &  13.764  &  3.24$\, \pm \,  $0.08  &  3.33$\, \pm \,  $0.06  &  0.40$\, \pm \,  $0.14  &  0.20$\, \pm \,  $0.10  &  1.03$\, \pm \,  $0.01  &  1.08$\, \pm \,  $0.06  \\ 
3C225B  &  1  &  145.565  &  13.764  &  3.24$\, \pm \,  $0.08  &  3.33$\, \pm \,  $0.06  &  0.40$\, \pm \,  $0.15  &  0.20$\, \pm \,  $0.10  &  1.03$\, \pm \,  $0.01  &  1.08$\, \pm \,  $0.06  \\ 
3C228.0  &  2  &  148.275  &  14.330  &  2.49$\, \pm \,  $0.13  &  2.52$\, \pm \,  $0.13  &  0.12$\, \pm \,  $0.05  &  0.04$\, \pm \,  $0.02  &  1.01$\, \pm \,  $0.00  &  1.00$\, \pm \,  $0.02  \\ 
3C234  &  2  &  150.454  &  28.787  &  1.55$\, \pm \,  $0.21  &  1.55$\, \pm \,  $0.21  &  0.03$\, \pm \,  $0.02  &  0.01$\, \pm \,  $0.01  &  1.00$\, \pm \,  $0.00  &  1.00$\, \pm \,  $0.02  \\ 
3C236  &  1  &  151.507  &  34.903  &  1.02$\, \pm \,  $0.33  &  0.93$\, \pm \,  $0.27  &  0.00$\, \pm \,  $0.00  &  0.01$\, \pm \,  $0.01  &  1.00$\, \pm \,  $0.00  &  1.00$\, \pm \,  $0.02  \\ 
3C237  &  1  &  152.000  &  7.505  &  1.64$\, \pm \,  $0.27  &  1.65$\, \pm \,  $0.23  &  0.34$\, \pm \,  $0.13  &  0.01$\, \pm \,  $0.02  &  1.01$\, \pm \,  $0.01  &  1.00$\, \pm \,  $0.02  \\ 
3C245B  &  1  &  160.684  &  12.059  &  2.14$\, \pm \,  $0.11  &  2.14$\, \pm \,  $0.10  &  0.01$\, \pm \,  $0.02  &  0.01$\, \pm \,  $0.01  &  1.00$\, \pm \,  $0.00  &  1.00$\, \pm \,  $0.02  \\ 
3C245A  &  1  &  160.686  &  12.059  &  2.14$\, \pm \,  $0.11  &  2.14$\, \pm \,  $0.10  &  0.00$\, \pm \,  $0.01  &  0.01$\, \pm \,  $0.01  &  1.00$\, \pm \,  $0.00  &  1.00$\, \pm \,  $0.02  \\ 
P1055+20  &  2  &  164.574  &  19.866  &  1.69$\, \pm \,  $0.07  &  1.70$\, \pm \,  $0.07  &  0.08$\, \pm \,  $0.05  &  0.01$\, \pm \,  $0.02  &  1.01$\, \pm \,  $0.00  &  1.00$\, \pm \,  $0.02  \\ 
1055+018  &  1  &  164.623  &  1.566  &  3.33$\, \pm \,  $0.25  &  3.25$\, \pm \,  $0.16  &  0.00$\, \pm \,  $0.00  &  0.01$\, \pm \,  $0.02  &  1.00$\, \pm \,  $0.00  &  1.00$\, \pm \,  $0.02  \\ 
P1117+14  &  2  &  170.810  &  14.318  &  2.20$\, \pm \,  $0.11  &  2.20$\, \pm \,  $0.11  &  0.00$\, \pm \,  $0.01  &  0.01$\, \pm \,  $0.01  &  1.00$\, \pm \,  $0.00  &  1.00$\, \pm \,  $0.02  \\ 
3C263.1  &  1  &  175.855  &  22.116  &  1.85$\, \pm \,  $0.15  &  1.85$\, \pm \,  $0.14  &  0.00$\, \pm \,  $0.01  &  0.01$\, \pm \,  $0.01  &  1.00$\, \pm \,  $0.00  &  1.00$\, \pm \,  $0.02  \\ 
3C264.0  &  2  &  177.019  &  19.592  &  2.84$\, \pm \,  $0.04  &  2.84$\, \pm \,  $0.04  &  0.02$\, \pm \,  $0.02  &  0.01$\, \pm \,  $0.01  &  1.00$\, \pm \,  $0.00  &  1.00$\, \pm \,  $0.02  \\ 
3C267.0  &  2  &  178.181  &  12.751  &  2.38$\, \pm \,  $0.07  &  2.39$\, \pm \,  $0.07  &  0.06$\, \pm \,  $0.04  &  0.01$\, \pm \,  $0.01  &  1.00$\, \pm \,  $0.00  &  1.00$\, \pm \,  $0.02  \\ 
3C272.1  &  2  &  186.960  &  12.846  &  1.94$\, \pm \,  $0.06  &  1.95$\, \pm \,  $0.06  &  0.03$\, \pm \,  $0.03  &  0.01$\, \pm \,  $0.01  &  1.00$\, \pm \,  $0.00  &  1.00$\, \pm \,  $0.02  \\ 
3C273  &  1  &  187.275  &  2.052  &  0.97$\, \pm \,  $0.71  &  0.98$\, \pm \,  $0.64  &  0.04$\, \pm \,  $0.05  &  0.01$\, \pm \,  $0.01  &  1.00$\, \pm \,  $0.00  &  1.00$\, \pm \,  $0.02  \\ 
3C274.1  &  2  &  188.862  &  21.343  &  2.49$\, \pm \,  $0.10  &  2.52$\, \pm \,  $0.10  &  0.12$\, \pm \,  $0.05  &  0.09$\, \pm \,  $0.04  &  1.01$\, \pm \,  $0.00  &  1.02$\, \pm \,  $0.02  \\ 
4C07.32  &  2  &  199.764  &  7.004  &  2.55$\, \pm \,  $0.11  &  2.61$\, \pm \,  $0.11  &  0.15$\, \pm \,  $0.08  &  0.09$\, \pm \,  $0.03  &  1.02$\, \pm \,  $0.00  &  1.02$\, \pm \,  $0.02  \\ 
4C32.44  &  1  &  201.569  &  31.903  &  1.11$\, \pm \,  $0.12  &  1.11$\, \pm \,  $0.10  &  0.03$\, \pm \,  $0.02  &  0.01$\, \pm \,  $0.02  &  1.00$\, \pm \,  $0.00  &  1.00$\, \pm \,  $0.02  \\ 
4C25.43  &  1  &  202.657  &  25.153  &  0.87$\, \pm \,  $0.14  &  0.86$\, \pm \,  $0.12  &  0.00$\, \pm \,  $0.00  &  0.01$\, \pm \,  $0.01  &  1.00$\, \pm \,  $0.00  &  1.00$\, \pm \,  $0.02  \\ 
3C286  &  1  &  202.785  &  30.509  &  0.84$\, \pm \,  $0.29  &  0.84$\, \pm \,  $0.27  &  0.05$\, \pm \,  $0.03  &  0.01$\, \pm \,  $0.01  &  1.00$\, \pm \,  $0.01  &  1.00$\, \pm \,  $0.02  \\ 
4C12.50  &  1  &  206.889  &  12.290  &  2.00$\, \pm \,  $0.10  &  2.02$\, \pm \,  $0.09  &  0.11$\, \pm \,  $0.05  &  0.07$\, \pm \,  $0.03  &  1.01$\, \pm \,  $0.01  &  1.01$\, \pm \,  $0.02  \\ 
3C293  &  2  &  208.073  &  31.449  &  1.09$\, \pm \,  $0.07  &  1.09$\, \pm \,  $0.07  &  0.05$\, \pm \,  $0.12  &  0.01$\, \pm \,  $0.01  &  1.00$\, \pm \,  $0.00  &  1.00$\, \pm \,  $0.02  \\ 
3C298  &  1  &  214.784  &  6.476  &  1.90$\, \pm \,  $0.13  &  1.91$\, \pm \,  $0.12  &  0.01$\, \pm \,  $0.01  &  0.02$\, \pm \,  $0.02  &  1.00$\, \pm \,  $0.00  &  1.00$\, \pm \,  $0.02  \\ 
4C20.33  &  2  &  216.958  &  19.997  &  2.14$\, \pm \,  $0.10  &  2.16$\, \pm \,  $0.10  &  0.13$\, \pm \,  $0.07  &  0.02$\, \pm \,  $0.02  &  1.01$\, \pm \,  $0.00  &  1.00$\, \pm \,  $0.02  \\ 
3C310  &  2  &  226.243  &  26.017  &  2.97$\, \pm \,  $0.15  &  3.18$\, \pm \,  $0.15  &  0.42$\, \pm \,  $0.16  &  0.13$\, \pm \,  $0.03  &  1.07$\, \pm \,  $0.00  &  1.04$\, \pm \,  $0.03  \\ 
3C315  &  2  &  228.418  &  26.125  &  3.90$\, \pm \,  $0.13  &  4.40$\, \pm \,  $0.13  &  0.51$\, \pm \,  $0.18  &  0.17$\, \pm \,  $0.03  &  1.13$\, \pm \,  $0.00  &  1.08$\, \pm \,  $0.03  \\ 
UGC09799  &  1  &  229.185  &  7.022  &  2.71$\, \pm \,  $0.10  &  2.72$\, \pm \,  $0.08  &  0.03$\, \pm \,  $0.02  &  0.03$\, \pm \,  $0.03  &  1.01$\, \pm \,  $0.00  &  1.00$\, \pm \,  $0.02  \\ 
4C04.51  &  1  &  230.310  &  4.506  &  3.85$\, \pm \,  $0.11  &  3.89$\, \pm \,  $0.10  &  0.05$\, \pm \,  $0.02  &  0.08$\, \pm \,  $0.02  &  1.01$\, \pm \,  $0.00  &  1.01$\, \pm \,  $0.02  \\ 
3C327.1A  &  1  &  241.187  &  1.298  &  6.76$\, \pm \,  $0.16  &  7.33$\, \pm \,  $0.10  &  0.26$\, \pm \,  $0.10  &  0.22$\, \pm \,  $0.06  &  1.09$\, \pm \,  $0.01  &  1.12$\, \pm \,  $0.04  \\ 
3C327.1B  &  1  &  241.190  &  1.297  &  6.76$\, \pm \,  $0.16  &  7.30$\, \pm \,  $0.10  &  0.25$\, \pm \,  $0.09  &  0.22$\, \pm \,  $0.06  &  1.08$\, \pm \,  $0.01  &  1.12$\, \pm \,  $0.04  \\ 
PKS1607  &  1  &  242.306  &  26.691  &  3.34$\, \pm \,  $0.11  &  3.41$\, \pm \,  $0.09  &  0.22$\, \pm \,  $0.08  &  0.08$\, \pm \,  $0.03  &  1.02$\, \pm \,  $0.01  &  1.01$\, \pm \,  $0.02  \\ 
J1613  &  1  &  243.421  &  34.213  &  1.37$\, \pm \,  $0.15  &  1.36$\, \pm \,  $0.10  &  0.00$\, \pm \,  $0.00  &  0.01$\, \pm \,  $0.01  &  1.00$\, \pm \,  $0.00  &  1.00$\, \pm \,  $0.02  \\ 
3C346  &  1  &  250.953  &  17.264  &  4.55$\, \pm \,  $0.07  &  4.73$\, \pm \,  $0.06  &  0.21$\, \pm \,  $0.08  &  0.21$\, \pm \,  $0.04  &  1.04$\, \pm \,  $0.00  &  1.09$\, \pm \,  $0.04  \\ 
PKS2127  &  1  &  322.637  &  5.038  &  4.50$\, \pm \,  $0.14  &  4.53$\, \pm \,  $0.10  &  0.08$\, \pm \,  $0.04  &  0.12$\, \pm \,  $0.05  &  1.01$\, \pm \,  $0.01  &  1.04$\, \pm \,  $0.04  \\ 
J2232  &  1  &  338.152  &  11.731  &  4.55$\, \pm \,  $0.16  &  4.64$\, \pm \,  $0.12  &  0.19$\, \pm \,  $0.07  &  0.08$\, \pm \,  $0.03  &  1.02$\, \pm \,  $0.01  &  1.01$\, \pm \,  $0.03  \\ 
3C454.3  &  1  &  343.492  &  16.148  &  5.95$\, \pm \,  $0.17  &  6.03$\, \pm \,  $0.11  &  0.26$\, \pm \,  $0.10  &  0.09$\, \pm \,  $0.03  &  1.01$\, \pm \,  $0.01  &  1.01$\, \pm \,  $0.03  \\ 
3C454.0  &  2  &  343.600  &  18.893  &  4.60$\, \pm \,  $0.13  &  4.65$\, \pm \,  $0.13  &  0.10$\, \pm \,  $0.04  &  0.07$\, \pm \,  $0.03  &  1.01$\, \pm \,  $0.00  &  1.01$\, \pm \,  $0.02  \\ 
3C459  &  1  &  349.147  &  4.088  &  4.95$\, \pm \,  $0.07  &  5.10$\, \pm \,  $0.06  &  0.18$\, \pm \,  $0.07  &  0.12$\, \pm \,  $0.03  &  1.03$\, \pm \,  $0.01  &  1.03$\, \pm \,  $0.03  \\ 
\hline   
\caption{Parameters for the \nabs\ \thi\ sightlines used for verifying the CNN model. (1): Source name; (2) Reference (1=\citet{murray2018b}; 2=\citet{heiles2003}); (3-4): RA, Dec coordinates; (5): \hi\ column density in the optically-thin limit (Equation~\ref{e:nhi_thin}); (6) total \hi\ column density  (Equation~\ref{e:nhi_iso}; See Section~\ref{sec:data}); (7) \fcnm, the CNM fraction observed (Equation~\ref{e:fcnm}); (8) \fcnm\ predicted by the CNN; (9) \rhi, the correction to the optically-thin \hi\ column density for $21\rm\,cm$ optical depth (Equation~\ref{e:rhi}); (10) \rhi\ predicted by the CNN. The full table will be available in the online journal.}
\label{tab:sources}

\end{longtable*}

\end{document}